\documentclass[aps,prl,twocolumn,amsmath,nofootinbib,longbibliography,amssymb,superscriptaddress,10pt]{revtex4-1}

\setcitestyle{super}
\usepackage[english]{babel}
\usepackage{graphicx}
\usepackage{ulem}
\usepackage{soul}
\usepackage{float} %added by Evyatar

\usepackage[svgnames]{xcolor} %added by Evyatar
\usepackage{comment}
\usepackage[hidelinks,breaklinks=true]{hyperref}
\usepackage{verbatim}
\usepackage{esint}
\usepackage{marginnote}
\usepackage{enumitem}
\hypersetup{colorlinks,allcolors=black}

\renewcommand{\vec}[1]{\boldsymbol{#1}}
\newcommand{\RNum}[1]{\uppercase\expandafter{\romannumeral #1\relax}}

\def \G {{\cal{G}}}

\def \beq {\begin{eqnarray}}
\def \eeq {\end{eqnarray}}

\begin{document}

%\title{A criterion for strangeness in disordered metals}

%\title{Signatures of strangeness in the Lorentz ratio of non-Fermi liquids}

%\title{A Criterion for Strangeness in the Lorentz ratio of non-Fermi liquids}

\title{A Criterion for Strange Metallicity in the Lorenz Ratio}

%\title{Do non-Fermi liquids obey the Weidemann-Franz law?}

\author{Evyatar Tulipman}
\email{Corresponding author: evyatar.tulipman@weizmann.ac.il}
\affiliation{Department of Condensed Matter Physics, Weizmann Institute of Science, Rehovot 76100, Israel}

\author{Erez Berg}
\affiliation{Department of Condensed Matter Physics, Weizmann Institute of Science, Rehovot 76100, Israel}

\date{\today}

\begin{abstract}
The Wiedemann-Franz (WF) law, stating that the Lorenz ratio $L = \kappa/(T\sigma)$ between the thermal and electrical conductivities in a metal approaches a universal constant $L_0=\pi^2 k_B^2/ (3 e^2)$ at low temperatures, is often interpreted as a signature of fermionic Landau quasi-particles. In contrast, we show that various models of weakly disordered non-Fermi liquids also obey the WF law at $T \to 0$. Instead, we propose using the leading low-temperature correction to the WF law, $L(T)-L_0$ (proportional to the inelastic scattering rate), to distinguish different types of strange metals. As an example, we demonstrate that in a solvable model of a marginal Fermi liquid, $L(T)-L_0\propto -T$. Using the quantum Boltzmann equation (QBE) approach, we find analogous behavior in a class of marginal- and non-Fermi liquids with a weakly momentum-dependent inelastic scattering. In contrast, in a Fermi liquid, $L(T)-L_0$ is proportional to $-T^2$. 
This holds even when the resistivity grows linearly with $T$, due to $T-$linear quasi-elastic scattering (as in the case of electron-phonon scattering at temperatures above the Debye frequency). Finally, by exploiting the QBE approach, we demonstrate that the transverse Lorenz ratio, $L_{xy} = \kappa_{xy}/(T\sigma_{xy})$, exhibits the same behavior.
\end{abstract}

\maketitle

%\tableofcontents

% \section{Introduction}

\section{Introduction}The properties of the anomalous normal state of high-$T_c$ superconductors and other %modern 
quantum materials, commonly dubbed `strange metals,' are one of the most elusive mysteries in condensed matter physics \cite{varma_colloquium_2020,hartnoll_planckian_2021}. In particular, despite myriad works, it is still unclear to what extent the underlying physics of such systems departs from Landau's Fermi-liquid (FL) paradigm and necessitates a non-FL (NFL) description. 

One of the hallmark characteristics of strange metals is the $T$--linear resistivity at extremely low temperatures. This behavior has been empirically linked with the notion of Planckian dissipation \cite{bruin_similarity_2013,cao_strange_2020,hartnoll_planckian_2021,legros_universal_2019,grissonnanche_linear-temperature_2021}, showing a %remarkable 
degree of universality throughout different experimental setups and hinting towards a strongly-correlated NFL nature for these systems. Albeit at odds with standard FL theory, $T$--linear resistivity can appear in FLs in the presence of certain scattering mechanisms, at least in some intermediate- to low-$T$ window \cite{das_sarma_charged_1999,das_sarma_screening_2015,wu_phonon-induced_2019,mousatov_theory_2020}. %or even from standard mechanisms in certain circumstances, at least in some intermediate- to low-$T$ window \cite{das_sarma_charged_1999,das_sarma_screening_2015,wu_phonon-induced_2019}. 
%In other words, while some aspects of strange metals appear universal, its origins might not be. Hence, 
It is thus crucial to develop ways to identify the mechanism of $T$--linear resistivity in strange metals.
%An experimental way to identify the %FL or NFL nature of different setups showing
%the mechanism of $T$--linear resistivity at low temperatures is an essential step toward 
%realistic theories of strange metals. 

Here we present a simple criterion for weakly disordered metals that sharply distinguishes different sources of $T$--linear resistivity. 
%By weakly disordered metals, we mean metallic systems where the resistivity is dominated by the residual resistivity due to impurity scattering as $T\to 0$. 
%To this end, we strictly ignore superconducting transitions, localization (or anti-localization) transitions or any other normal state instabilities that may arise in the $T\to 0$ limit. 
Our criterion is based on the behavior of the low-$T$ leading correction to the Lorenz ratio, $L(T) = \frac{\kappa}{T\sigma}$, with $\kappa$ and $\sigma$ being the thermal and electrical conductivities, respectively. %We shall first recall the WF law relating $\sigma$, and then describe our criterion. 

The Weidemann-Franz (WF) law \cite{franz_ueber_1853} states that %is an empirical law that 
%relates the thermal and electrical conductivities, such that
\begin{eqnarray}
 \overline{L}\left(T\right) \equiv \frac{L}{L_0} \to 1 
 \label{wf_function}
\end{eqnarray}
as $T \to 0$. Here, $L_0 = \pi^2 k_B^2/ (3 e^2)$ is the so-called Lorenz number (we set $e=k_B = 1$ henceforth). 
Roughly speaking, the deviation of $\overline{L}(T)$ from 1 serves as a measure for the relative contribution of inelastic scattering to charge and thermal transport ($\overline{L}(T)\approx1$ implies that elastic or quasi-elastic scattering is dominant) \cite{Ziman_2001}. Dominantly inelastic scattering leads to deviations from the WF law in many circumstances \cite{mahajan_non-fermi_2013,lavasani_wiedemann-franz_2019,stangier_breakdown_2022,principi_violation_2015}. 

\begin{figure}[t]
\centering

\includegraphics[width=\columnwidth]{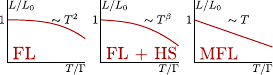}

\caption{
Schematic plot of the low-$T$ behavior of the normalized Lorenz ratio % as a function of $T$ %in the limit $T/\Gamma \to 0$ 
for %weakly disordered 
systems that obey the Wiedemann-Franz law at $T=0$. Here, $T$ is assumed to be smaller than $\Gamma$, the elastic scattering rate. The $T$ dependence of the leading deviation from $L=L_0$ serves as a criterion for strange metallicity: Fermi liquids (FL) exhibit $L/L_0 - 1 \propto -T^2$; Fermi liquids with hot spots (FL+HS) are characterized by $L/L_0 - 1 \propto - T^\beta$, $1<\beta <2$; and certain marginal Fermi-liquids (MFL) have $L/L_0 - 1\propto -T$.} 
\label{fig:schematic_wfl}
\end{figure}

The validity of the WF law is often used as a test for the existence of FL-like quasi-particle excitations at the lowest temperatures \cite{proust_heat_2002,Makariy2007,reid_wiedemann-franz_2014,grissonnanche_wiedemann-franz_2016,michon_wiedemann-franz_2018}. %is not generally true, namely, 
However, the fact that WF is obeyed does not necessarily imply that transport is carried by FL quasi-particles \cite{schwiete_thermal_2014,wang2022wiedemannfranz,Ulaga2022}. Indeed, as we shall show, one can construct solvable models of NFLs where the WF law is obeyed at $T\to 0$. 
%Instead, we argue that the proper litmus test for the existence of FL quasiparticles is the behavior of the leading order deviation from the WF law at low $T$. 
%In the context of strange metallicity, 
The known mechanisms for $T$--linear resistivity (not necessarily extending down to $T\to 0$) in FLs involve elastic or quasi-elastic scattering. These include electron-phonon scattering \cite{wu_phonon-induced_2019} or static charged impurities in 2D \cite{das_sarma_charged_1999,das_sarma_screening_2015}. In contrast, $T$--linear resistivity associated with NFLs is typically associated with inelastic scattering \cite{chowdhury_sachdev-ye-kitaev_2022,lee_low-temperature_2021,patel_universal_2022,chowdhury_translationally_2018,patel_magnetotransport_2018}. In both cases, however, assuming that ($T$-independent) impurity scattering dominates at sufficiently low $T$, we expect the WF law to be obeyed at $T \to 0$. Hence, in order to learn about the FL or NFL origin of the $T$--linear resistivity, one must consider the leading low-temperature deviation from the WF law (see Fig.~\ref{fig:schematic_wfl}).

Our criterion is applicable to systems that obey the WF law
%The applicability of our criterion relies on the validity of the WF 
at $T\to 0$, as in the cuprates at sufficiently low temperature~\cite{proust_heat_2002,grissonnanche_wiedemann-franz_2016,michon_wiedemann-franz_2018}. In this context, it is worth noting that certain weakly disordered 2D systems with Coulomb interactions are expected to violate the WF law at $T \to 0$ \cite{Niven2005,schwiete_theory_2016,xie_transport_2016}. 
However, in metals, the deviation from the WF law is significant at an exponentially small temperature in $k_F \ell$, where $k_F$ is the Fermi momentum and $\ell$ is the elastic mean free path. Our discussion applies above this energy scale. 

\section{Results}

\subsection{A criterion for strangeness}
\label{criterion section}
Consider weakly disordered metals (in 2 or 3 spatial dimensions), such that the dc resistivity has the following form as $T\to 0$: $\rho\left(T\right) = \rho_0 + AT^\alpha$, where $\rho_0$ is the residual resistivity, and $A,\alpha>0$. %{\color{purple} We assume that the $T\to 0$ limit of electronic thermal resistivity obeys $\rho_{\rm th} \equiv T/\kappa_{\rm el} = \rho_{\rm th,0} + B T^{\beta}$.} % and $\alpha \geq 1$. 
We assume that impurity scattering dominates at sufficiently low $T$, and the WF law is satisfied at $T\to0$. In this case, the low-$T$ electronic thermal resistivity takes the form $\rho_{\rm th}(T)\equiv T/\kappa_{\rm el} = \rho_{\rm th,0} + B T^{\beta}$ with $B,\beta>0$. The normalized Lorenz ratio \eqref{wf_function} then takes the following form:
\begin{eqnarray}
\overline{L}\left(T\right) -1\propto - T^{\beta}.
\end{eqnarray}
We claim that the exponent $\beta$ is universal and provides information on the nature of the system. In ordinary FLs, 
%\footnote{Logarithmic corrections may arise due to electron-electron interactions in 2D \cite{Mishchenko2003,das_sarma_know_2021}, such that $\overline{L}= 1-c T^2 \log{\left(1/T\right)}$ is also consistent with FLs.}
$\beta = 2$ (logarithmic corrections may arise due to electron-electron interactions in 2D \cite{Mishchenko2003,das_sarma_know_2021}). Systems where a portion of Fermi surface (FS) is `hot', while the rest is FL-like, have $1<\beta<2$. 
%typically indicate that a portion of Fermi surface (FS) is `hot', while the rest of the system is a FL. 
Most interestingly, if $\beta \leq 1$, the system is not described by any existing theory of a FL. In particular, the case $\alpha = \beta = 1$ arises in certain models that realize marginal Fermi-liquids (MFLs) \cite{varma_phenomenology_1989}. We therefore argue that $\alpha = \beta  = 1$ could serve as a criterion for `strangeness', in the sense that it signals a full departure from FL theory. See Fig.~\ref{fig:schematic_wfl} for a schematic illustration of the different cases.  

\label{criterion}

\subsection{Fermi liquids}We consider a weakly disordered FL with electron-electron (el-el) or electron-phonon (el-ph) interactions.
%, where the weak disorder.  
%is described by static (charged or neutral) isotropic impurities. 
We assume that the WF law is obeyed at $T\to 0$ due to the dominance of elastic scattering ~%of off static impurities 
\cite{Ziman_2001,sykes_transport_1970,michaeli_quantum_2009,lavasani_wiedemann-franz_2019,lucas_electronic_2018,mahajan_non-fermi_2013,castellani_thermal_1987}. Here, and in the following section, the disorder corresponds to static impurities, which provide a source of elastic scattering with rate $\Gamma$. 
%{\color{red} Note that the disorder corresponds to the presence of static impurities that provide a source of elastic scattering with rate .}
%In addition, we restrict ourselves the low temperatures, where in particular $T\ll E_{\rm F}$. Systems with vanishingly small densities, where violation of the WF law are known to occur \cite{xie_transport_2016-1}, are therefore beyond the scope of our discussion.  

At $T>0$, el-el and el-ph interactions provide inelastic scattering mechanisms that lead to deviations from the WF law. The contribution of el-el interactions, a hallmark of FL theory, lead to resistivities of the form $\rho = \rho_0 + AT^2$ (assuming Umklapp scattering is present) and $\rho_{\rm th} = T/\kappa = \rho_{\rm th,0} + BT^2$ (see e.g. \cite{abrikosov_theory_1959,sykes_transport_1970,hojgard_jensen_exact_1968,Paglione2005}), which translates to 
\begin{eqnarray}
\overline{L} \left( T \right)  -1\propto - T^2 
\label{FL_WF_deviation}
\end{eqnarray}
where the negative slope is related to the additional contribution of forward scattering that relaxes the thermal current, but not the electrical current \cite{Ziman_2001,mahajan_non-fermi_2013,lavasani_wiedemann-franz_2019}. The el-ph contribution to the electrical (thermal) resistivity is $\mathcal{O} \left( T^{d+2} \right)$ ($\mathcal{O} \left( T^{d} \right)$), respectively (where $d>1$), as long as $T$ is much smaller than $T_{\rm BG}$, the Bloch-Gruneisen temperature~\cite{Ziman_2001}. That is, the el-ph contribution is subleading in 3D, while in 2D it may modify the non-universal slope, such that the form \eqref{FL_WF_deviation} holds at sufficiently low $T$ in a FL.

In fact, Eq.~\eqref{FL_WF_deviation} %applies even for the two known examples of mechanisms for 
applies even in cases where the resistivity of a FL is $T$--linear. 
%, either due to scattering off impurity scattering in $2D$, or scattering off quasi-classical acoustic phonons. Let us begin with the 2D charged impurities case (in 3D, the contribution to the resistivity is $\mathcal{O} \left( T^2 \right)$) \cite{das_sarma_charged_1999,das_sarma_screening_2015,xie_transport_2016}. 
For example, Coulomb screening of charged impurities, treated within the random phase approximation, leads to a $T$--linear resistivity in a 2D FL, 
due to thermal suppression of the FL polarizability~\cite{Stern1980,das_sarma_screening_2015}. (In 3D, this contribution to the resistivity is $\mathcal{O} \left( T^2 \right)$ \cite{das_sarma_charged_1999,das_sarma_screening_2015,xie_transport_2016}.) 
%Importantly, this $T$--linear resistivity stems from a $T$-dependence of the screened impurity potential. 
%, namely, the residual resistivity due to elastic scattering off of impurities acquires a $T$ dependence. 
However, in this case, the $T-$linear scattering is still essentially elastic, and the deviations from the WF law still obey Eq.~\eqref{FL_WF_deviation}. 
%{\color{red} Is inelastic scattering (associated with the polarizability at non-zero frequency) important?} {\color{blue} not within RPA}%the assumption that 
%the WF law is obeyed if the sole scattering mechanism is elastic impurity scattering is true regardless of the $T$-dependence of the impurity potential. 
%That is, in this case, $\overline{L} \left( T \right) = 1$ even as the resistivity increases linearly with $T$. 
%The leading correction to $\overline{L} =1$ is thus due to el-el interactions, such that \eqref{FL_WF_deviation} holds as before. 
%The same argument is expected to hold if another intrinsic $T$-dependence of impurities is responsible for the $T$--linear resistivity as $T\to 0$ [steve?]. 

Unlike the case of charged impurities, $T$--linear resistivity from el-ph interactions emerges only at temperatures $T\gtrsim T_{\rm BG}$ \cite{Ziman_2001}. Hence, this mechanism is always irrelevant at the limit $T\to 0$. On a more practical note, if $T_{\rm BG}$ sets a particularly small energy scale, the $T$--linear resistivity due to el-ph scattering might appear to extend down to the lowest experimentally accessible temperatures (as long as $T \gtrsim T_{\rm BG}$). However, 
%this $T$--regime {\color{blue} corresponds to the 
in this ``equipartition'' regime, phonons are essentially classical and the el-ph scattering is quasi-elastic. Hence, the WF law is essentially obeyed in this regime \cite{Ziman_2001}.  
%Hence, similarly to the previous case, {\color{blue} the equipartition} %this $T$--
%regime shows $T$--linear resistivity and $\overline{L} = 1$. {\color{red} Something is not right here. $\overline{L}$ is the deviation from WF. Presumably $\overline{L} \approx  0$?} 
%, as expected from elastic scattering, or from quasi-elastic scattering due to phonons at high temperature.   

%\section{Fermi surfaces with hot spots}

%\label{HSFL_section}
\subsection{Fermi surfaces with hot spots}We now consider systems where a portion of the Fermi surface becomes `hot',  i.e., it experiences enhanced scattering with an anomalous $T$--scaling. In some situations, such `hot spots' can lead to an anomalous $T$ dependence of the transport coefficients. This situation arises either when the system is on the verge of a finite wavevector instability~\cite{hlubina_resistivity_1995,hlubina_effect_1996,rosch_interplay_1999,syzranov_conductivity_2012,hartnoll_quantum_2011,herman_deviation_2019}, or when the system is turned to a Van Hove singularity where the topology of the Fermi surface changes~\cite{mousatov_theory_2020,stangier_breakdown_2022}. 
%In some systems, the vicinity of various instabilities may lead to anomalous $T$-scaling of transport coefficients, deviating from conventional FL theory. This typically arises in situations where a portion of the FS becomes `hot', i.e., experiences enhanced scattering from fluctuations of the nearby instability \cite{hlubina_resistivity_1995,hlubina_effect_1996,rosch_interplay_1999,syzranov_conductivity_2012,hartnoll_quantum_2011,herman_deviation_2019,stangier_breakdown_2022}. In what follows we demonstrate the validity of our criterion in two tractable examples with $\boldsymbol{Q}=0$ and $\boldsymbol{Q}\ne 0$ instabilities, where $\boldsymbol{Q}$ is the ordering wave-vector associated with the relevant instability. 

Consider the low--$T$ behavior of $\overline{L}\left( T \right)$ in a 2D system where a Van Hove singularity (VHS) crosses the FS in the vicinity of a %strain-induced 
Lifshitz transition \cite{mousatov_theory_2020, stangier_breakdown_2022}. %This case can be thought of as a $\boldsymbol{Q}=0$ instability, in the sense that 
In this case, we refer to the Fermi surface regions near the VHS as `hot'. The transport scattering rates are dominated by processes where a `cold' electron (away from the VHS) is scattered by a `hot' one, or two cold electrons are scattered and one of them ends up near the VHS. In clean systems, this leads to $\rho\sim T^2\log(1/T)$ \cite{hlubina_effect_1996, mousatov_theory_2020}   and $\rho_{\rm th}\sim T^{3/2}$~\cite{stangier_breakdown_2022}. 
%The Van Hove singularity leads to isolated `hot' regions in the FS with an enhanced scattering rate. %, as opposed to the hot lines connected by the AFM wavevector we considered above. 
%In particular, unlike the previous example, where transport properties are dominated by the hot lines at sufficiently low $T$, the anomalous transport coefficients here are attributed to scattering events that combine hot spots and generic, cold areas of the FS. 
This behavior persists in the presence of impurities, namely, $\rho = \rho_0 + AT^{2}\log(1/T)$ and
$\rho_{\rm th} = \rho_{\rm th,0} + BT^{3/2}$~\cite{hlubina_effect_1996, stangier_breakdown_2022}, such that the deviation from WF law satisfy
\begin{eqnarray}
\overline{L} \left( T \right)  -1\propto - T^{3/2}.
\label{hot_FL_WF_deviation}
\end{eqnarray}
%{\color{red} This result is actually not derived in Stangier et al. Is it obvious that the correction to the residual electrical resistivity scales as $T^{3/2}$? {\color{blue} I actually took this result directly from the Stangier et al. paper.} Is it because the ch$\to$ch $T^{3/2}$ rate is averaged over the entire FS?} {\color{blue} Yes, to the best of my understanding. But I didn't find any explicit calculations for this case (in the paper or related references).}

We proceed by considering a weakly disordered FL near an antiferromagnetic (AFM) quantum critical point in 3D, as studied in Refs.~\cite{rosch_interplay_1999,syzranov_conductivity_2012}. In this case, the FS contains `hot lines' connected by the non-zero AFM wavevector, where the scattering off spin fluctuation is most effective. The hot lines then acquire anomalous, NFL-like, scattering rates which may manifest in transport coefficients. In the absence of impurities, these hot lines are short-circuited by the remaining `cold' parts of the FS such that transport coefficients follow the conventional FL behavior at sufficiently low $T$ \cite{hlubina_resistivity_1995}. However, introducing impurities enables the hot lines to participate in transport, since, loosely speaking, the scattering rate is averaged over the entire FS. Ref.~\cite{rosch_interplay_1999} showed that this leads to an anomalous $T$--scaling of the resistivity, where $\rho = \rho_0 + AT^{3/2}$ at the lowest temperatures. By extending the analysis of \cite{rosch_interplay_1999} to the thermal conductivity, we find that the thermal resistivity follows the same anomalous behavior: $\rho_{\rm th} = \rho_{\rm th,0} + BT^{3/2}$, see Supplementary Material. Combining the two resistivities, the deviation from WF law follows Eq.~\eqref{hot_FL_WF_deviation}. 
%leads to  %{\color{red} Do we want to comment on the 2D case as well?} {\color{blue} I think not, if we want to restrict ourselves to tractable cases. The 2D case is still quite controversial, isn't it? } {\color{red} It's not entirely solved, but I think we should nevertheless say something about it. At least we can say that it is not entirely solved.} {\color{blue} added a short comment below}

Interestingly, a straightforward generalization of the analysis above to 2D yields $\rho = \rho_0 + AT$ \cite{syzranov_conductivity_2012}. The same reasoning is expected to hold for the thermal resistivity, which would imply that $\overline{L} -1 \propto -T$ in 2D. However, this analysis is based on the Hertz-Millis treatment of the AFM QCP, which breaks down at sufficiently low temperatures in the 2D case \cite{Abanov2004,lohneysen_fermi-liquid_2007}.  

%\section{Marginal Fermi liquids}

%\label{mfl_section}

\subsection{Marginal Fermi liquids}In this section, we construct a solvable model of a 2D weakly disordered MFL that shows $T$--linear resistivity down to the lowest temperatures and obeys the WF law at $T \to 0$, with a leading correction of the form
\begin{eqnarray}
\overline{L} \left( T \right) -1 \propto - T. 
\label{MFL_WF_deviation}
\end{eqnarray}
In addition, we comment on the expected behavior of other tractable models of MFLs in 2 and 3 dimensions, suggesting that Eq.~\eqref{MFL_WF_deviation} could be a robust signature of a class of weakly disordered MFLs. We further corroborate this expectation using the Quantum Boltzmann Equation (QBE) approach in the following section.

Consider a weakly disordered variant of the model studied in Ref.~\cite{chowdhury_translationally_2018}, based on a 2-band lattice generalization of the Sachdev-Ye-Kitaev (SYK) model~\cite{sachdev_gapless_1993,Kitaev_SYK_talk,maldacena_remarks_2016}. The model is defined on a D--dimensional lattice, and contains two species of fermions, $\{c\}$ and $\{f\}$, each containing $N$ orbitals per unit cell, governed by the Hamiltonian $H=H_{c}+H_{f}+H_{cf}$, 
 where 
\begin{eqnarray}
H_{c}&=&-\sum_{\boldsymbol{r},\boldsymbol{r}',l}\left(t_{\boldsymbol{r},\boldsymbol{r}'}+\mu\delta_{\boldsymbol{r},\boldsymbol{r}'}\right)c_{\boldsymbol{r}l}^{\dagger}c_{\boldsymbol{r}'l}+\frac{1}{N^{1/2}}\sum_{\boldsymbol{r},ij}W_{ij\boldsymbol{r}}c_{\boldsymbol{r}i}^{\dagger}c_{\boldsymbol{r}j} \nonumber ;\\ 
H_{cf}&=&\frac{1}{N^{3/2}}\sum_{\boldsymbol{r,r}'}\sum_{ijkl}V_{ijkl}\Upsilon_{\boldsymbol{r,r}'} c_{\boldsymbol{r}i}^{\dagger}f_{\boldsymbol{r}'j}^{\dagger}c_{\boldsymbol{r}k}f_{\boldsymbol{r}'l}  ; \\
H_{f}&=&\frac{2}{N^{3/2}}\sum_{ ijkl}U_{ijkl} f_{\boldsymbol{r}i}^{\dagger}f_{\boldsymbol{r}j}^{\dagger}f_{\boldsymbol{r}k}f_{\boldsymbol{r}l}. \nonumber \label{MFL_SYK_H}
\end{eqnarray}
The hopping matrix $t_{\boldsymbol{r},\boldsymbol{r}'}$ is diagonal in orbital space and depends only on the distance $\left|\boldsymbol{r}-\boldsymbol{r}'\right|$. The last term in $H_c$ describes on-site disorder for the $c$-fermions, where $W_{ij \boldsymbol{r}}$ are site-dependent Gaussian random independent potential, satisfying $\overline{W_{ij\boldsymbol{r}}}=0,\overline{W_{ij\boldsymbol{r}}W_{ij\boldsymbol{r}'}}=W^2\delta_{\boldsymbol{r},\boldsymbol{r}'}$.  
%\delta_{\{i,j\},\{l,m\} }$.  
The couplings in $H_{cf}$ and $H_f$ are site-independent Gaussian random independent variables, satisfying $\overline{V_{ijkl}}=0,\overline{V^2_{ijkl}}=U_{cf}^2$ and similarly for $U_{ijkl}$ (with variance $U_f^2$). The function $\Upsilon$ determines the spatial dependence of the $cf$-interaction. Note that for $W=0$, the model is translationally invariant for every realization of the interactions. We first consider the case of onsite interaction as in \cite{chowdhury_translationally_2018}: $\Upsilon_{\boldsymbol{r,r}'} = \delta_{\boldsymbol{r,r}'}$. Spatially extended $\Upsilon$ will be considered later on. 

The model is solvable in the $N\to \infty$ limit, where its properties are dictated by replica-diagonal saddle-point of the real- and imaginary-time effective action~\cite{chowdhury_translationally_2018}. The low-energy saddle-point equations describe SYK-like, incoherent $f$-fermions. These $f$-fermions constitute a local quantum critical bath for the $c$-fermions, giving rise to a weakly disordered MFL form for the Green's function of the $c$-fermions. Importantly, the onsite disorder $W$ for the $c$-fermions does not alter the low-energy behavior of the $f$-fermions, rather it only enters as an additional $T$-independent, elastic scattering term to the $c$-fermions. For example, at $T=0$, the Matsubara frequency Green's function is of the form
\begin{eqnarray}
G_{c}\left(\boldsymbol{k},i\omega\right)&=&\frac{1}{i\omega-\varepsilon_{\boldsymbol{k}}-\Sigma_{c}\left(i\omega\right)};\\
\Sigma_{c}\left(i\omega\right) &=& -i\frac{\Gamma}{2}\text{sgn}\left(\omega\right)+ \Sigma_{cf}\left(i\omega\right);\\
\Sigma_{cf}\left(i\omega\right)&=&-\frac{\nu_{0}U_{cf}^{2}}{2\pi^{2}U_{f}}i\omega\log\left(\frac{U_{f}}{\left|\omega\right|}\right),
\end{eqnarray}
where $\Gamma = 2\pi \nu_0 W^2$ is the disorder energy scale. 

\begin{figure}[t]
\centering

\includegraphics[width=\columnwidth]{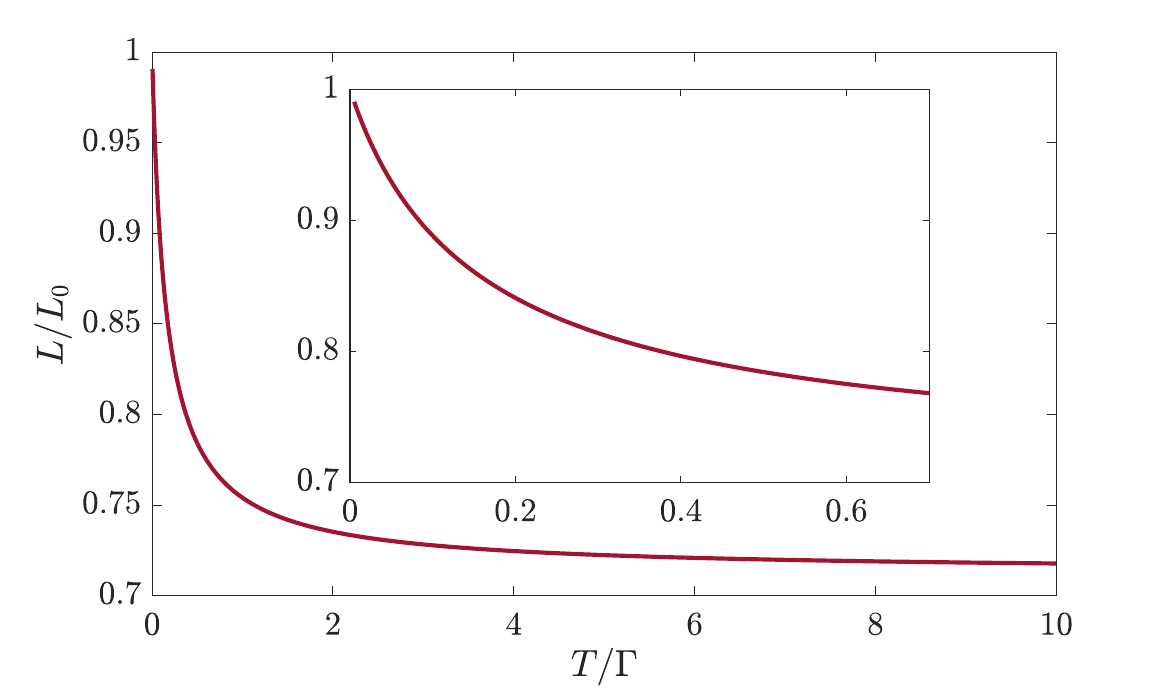}

\caption{ The Lorenz ratio as a function of temperature for the MFL model \eqref{MFL_SYK_H} with local $cf$-interaction. $\Gamma$ is the elastic scattering rate. The inset shows the limit $T\ll\Gamma$ which obeys Eq.~\eqref{MFL_WF_deviation}.} 
\label{fig:current}
\end{figure}

We proceed to consider transport. We compute the electrical and thermal conductivities using the Kubo formula. By virtue of the locality of the $f$-fermions, both conductivities are given in terms of the bare bubble expressions, similarly to Refs.~\cite{chowdhury_translationally_2018,patel_magnetotransport_2018}. We obtain the thermal conductivity, 
\begin{eqnarray}
\kappa=\frac{v_{F}^{2}\nu_{0}}{16 T^2}\int\frac{d\epsilon}{2\pi}\frac{\epsilon^{2}}{\left|\Sigma''_R\left(\epsilon \right) \right|}\text{sech}^{2}\left(\frac{\epsilon}{2T}\right)
\label{th_cond}
\end{eqnarray}
and the electrical conductivity,
\begin{eqnarray}
\sigma=\frac{ v_{F}^{2}\nu_{0}}{16 T}\int\frac{d\epsilon}{2\pi}\frac{1}{\left|\Sigma''_R\left(\epsilon\right)\right|}\text{sech}^{2}\left(\frac{\epsilon}{2T}\right).
\label{el_cond}
\end{eqnarray}
The imaginary part of the retarded self-energy is given by $-\Sigma''_R\left(\epsilon\right) = \frac{\Gamma}{2}+ g^2 {\rm Im} \left[ \epsilon \psi \left( \frac{-i\epsilon}{2\pi T}\right) + i \pi T \right]$ with $g^2 = \frac{U_{cf}^{2} \nu_0}{2 \pi^2 U_f}$ and where $\psi(z)$ is the digamma function \cite{patel_magnetotransport_2018}. Using Eq.~\eqref{th_cond} and Eq.~\eqref{el_cond}, we find that the WF law is obeyed at $T\to 0$, despite the fact that the MFL description of the $c$-fermions persists to the lowest temperatures, and that the leading deviation from the WF law obeys Eq.~\eqref{MFL_WF_deviation}. The Lorenz ratio $L(T)/L_0$ as a function of $T$ is shown in Fig.~\ref{fig:current}. As can be seen in the figure, $L/L_0$ decreases linearly with $T$ at small $T$, and saturates to the value correspodning to the clean case, $L/L_0\approx 0.71$~\cite{patel_magnetotransport_2018,maebashi_quantum-critical_2022}, at $T\gtrsim \Gamma$.

In order to examine the robustness of these results to details of the model, we consider the addition of spatially extended 
%a small nearest-neighbors 
$cf$-interactions: $\Upsilon_{\boldsymbol{r},\boldsymbol{r}'} = \delta_{\boldsymbol{r},\boldsymbol{r}'} + \eta \sum_{\boldsymbol{\boldsymbol{\delta}} = \pm \hat{x},\pm \hat{y}}  \delta_{\boldsymbol{r},\boldsymbol{r}'+\boldsymbol{\delta}}$ with $\eta$ being a small control parameter. This modification does not change the MFL form of the self-energy of the $c$-fermions. In addition, the form of the thermal current operator is unchanged, see Supplementary Material.
%since the $c$-fermions remain local in the $cf$-interaction, and the electrical current is unchanged since the $cf$-interaction preserves $U\left(1\right)$ locally. 
Hence, to leading order in $\eta$, the conductivities are given by $\alpha_{\eta} = \alpha_{0} + \delta \alpha$ for $\alpha = \sigma,\kappa$, where we have denoted $\alpha_{\eta = 0} \equiv \alpha_0$, and the correction $\delta \alpha $ is $ \mathcal{O}\left( \eta \right)$ and corresponds to the current bubble with an insertion of a single $cf$-interaction rung, see Supplementary Material. These corrections alter the Lorenz ratio, such that for $T\gg \Gamma$,
\begin{eqnarray}
L = \frac{\kappa_{0}}{T \sigma_{0}}\left( 1 + \frac{\delta \kappa}{\kappa_{0}} - \frac{\delta \sigma}{\sigma_{0}}\right) \ne \frac{\kappa_{0}}{T \sigma_{0}} ,
\label{non_universal_high_T_value}
\end{eqnarray}
which demonstrates that the saturation value is not universal. Importantly, the spatially-extended $cf$-interaction do not alter the $T \to 0$ behavior of the Lorenz ratio, which obeys Eq.~\eqref{MFL_WF_deviation}. We will demonstrate this and further highlight the conditions for which Eq.~\eqref{MFL_WF_deviation} is valid within the framework of the QBE in the next section.

It is worth commenting that the simplicity of the analysis of \eqref{MFL_SYK_H} comes with a price in the form of a residual $T\to 0$ extensive entropy due to the SYK-nature of the $f$-fermions~\cite{Kitaev_SYK_talk,maldacena_remarks_2016,chowdhury_translationally_2018,patel_magnetotransport_2018}. The residual entropy is relieved upon allowing quadratic terms in the $f$-fermions, but these also lead to 
%a recovery of 
FL behavior at low temperatures \cite{chowdhury_translationally_2018}. 
Nevertheless, we expect Eq.~\eqref{MFL_WF_deviation} to be a robust property of weakly disordered MFLs in 2 and 3 dimensions that show $T$--linear electrical resistivity, as we discuss in the next section. 

Let us briefly note that the results presented here and in the next section can be generalized to $f$-fermions governed by an SYK$_q$ ($q>4)$ Hamiltonian, while the $cf$-interaction is unchanged. For $q>4$, the $c$-fermions realize an incoherent, NFL description with $\rho = \rho_0 + AT^{4/q}$ and $\rho_{\rm th} = \rho_{\rm th,0} + BT^{4/q}$, such that $\overline{L} -1\propto -T^{4/q}$ \cite{chowdhury_translationally_2018}.
%Let us briefly note that the results presented in Sec.~\ref{mfl_section} can be generalized to $f$-fermions govenered by an SYK$_q$ ($q>4)$ Hamiltonian, which leads to incoherent $c$-fermions with $\overline{L} = 1-cT^{4/q}$ \cite{chowdhury_translationally_2018}. {\color{blue} NEW FOOTNOTE FOR SYK-q}

%\section{QBE approach}
%\label{QBE_section}

%The validity of the WF law at $T\to 0$ in MFLs may seem surprising at first. In conventional FL theory, the WF law follows from the fact that transport coefficients can be derived from a QBE that governs the Fermi distribution function of the electronic quasi-particles  \cite{Ziman_2001,mahajan_non-fermi_2013}. Interestingly, 
\subsection{Quantum Boltzmann equation approach}Even in the absence of well-defined quasi-particles, we may still derive a QBE for a generalized Fermi distribution function in the model of the previous section. Here we briefly outline the idea behind the QBE approach for MFLs and the conditions for which it is applicable. In addition, we highlight its implications on the validity of the WF law and the criterion for strangeness in a certain class of MFLs, using a generalization of the model \eqref{MFL_SYK_H} as a simple representative. We elaborate on several issues and supply technical details in the Supplementary material.

To derive a QBE in the absence of well-defined quasi-particles, we utilize the MFL form of the self-energy and the fact that the spectral function of the $c$-fermions is sharply peaked at the FS as a function of $\varepsilon_{\boldsymbol{k}}$ (this is in contrast to the QBE approach for FLs which relies on the sharp quasi-particle peak as a function of $\omega$). Within this approximation, known as the Prange-Kadanoff (PK) reduction scheme \cite{prange_transport_1964,nave_transport_2007}, the momenta of the $c$-fermions are restricted to the FS. Roughly speaking, the PK reduction is valid when the width of the electronic spectral function $\mathcal{A}(\omega\sim T,\boldsymbol{k})$ as a function of $\boldsymbol{k}$ is smaller than the typical momentum transfer in both elastic and inelastic scattering events, see Supplementary Material and ~\cite{guo_large_2022}. 
%is consistent in the presence of disorder provided that the inelastic scattering rate is weakly momentum dependent. Scattering events normal to the FS can therefore be neglected due to their small spectral weight \cite{prange_transport_1964,guo_large_2022}.

Considering the MFL model \eqref{MFL_SYK_H}, the QBE approach illustrates that 
\begin{enumerate}
    \item[(i)] The WF law may hold at $T \to 0$ due to the dominance of elastic scattering, regardless of the existence of well-defined quasi-particles;
    \item[(ii)] The leading deviation from the WF law obeys Eq.~\eqref{MFL_WF_deviation} in weakly disordered MFLs that admit the PK reduction scheme;
\end{enumerate}
%(i) the WF law may hold at $T \to 0$ due to the dominance of elastic scattering, regardless of the existence of well-defined quasi-particles; And (ii) the leading deviation from the WF law obeys Eq.~\eqref{MFL_WF_deviation}, 
where (ii) can be understood as a consequence of Matthiessen's rule. We further find that the deviation in Eq.~\eqref{MFL_WF_deviation} hold for a class of generalized models with spatially extended $cf$-interactions, see e.g., the previous section, which 
%reveals
confirms that (i) and (ii) have a much broader regime of validity in weakly disordered MFLs (and NFLs). Specifically, assuming that the momentum-dependence of the inelastic scattering rate is sufficiently weak (as defined above), such that PK reduction can be applied, the QBE approach suggests that WF law should hold at $T\to 0$. Moreover, since in these circumstances the transport relaxation rate are proportional to the single particle scattering rate, the leading low-$T$ deviation from the WF law is expected to satisfy Eq.~\eqref{MFL_WF_deviation}. 

\subsection{Transverse Lorenz ratio}

We employ the QBE approach to generalize our discussion to the transverse Lorenz ratio:
\begin{eqnarray}
    L_{xy} \equiv \frac{\kappa_{xy}}{T \sigma_{xy}},
\end{eqnarray}
where $\sigma_{xy}$ and $\kappa_{xy}$ denote the transverse electrical and thermal conductivities, respectively. Specifically, by solving the linearized QBE of the weakly disorder MFLs \eqref{MFL_SYK_H}, we find that the leading deviation from the (transverse) WF law for a class of MFLs follows the same scaling as the longitudinal: 
\begin{eqnarray}
    \overline{L}_{xy} - 1 \propto - T,
\end{eqnarray} 
as in Eq.~\eqref{MFL_WF_deviation}; see Supplementary Material. Moreover, while the derivation of the transverse conductivities is slightly more involved due to the presence of a weak magnetic field, the key ingredient remains the validity of the PK reduction scheme. This has the remarkable implication that, as long as the PK reduction scheme is valid, our conclusions for the longitudinal Lorenz ratio (i.e. (i) and (ii) from the previous section) equally apply to the transverse Lorenz ratio of weakly disordered MFLs (or NFLs). In addition, while the transverse conductivities are proportional to the applied magnetic field, this proportionality factor cancels in $L_{xy}$ such that the leading deviation is independent of the magnetic field. 

Note further that the extension of our criterion to the transverse Lorenz ratio holds also for weakly disordered FLs, where the leading deviation satisfies $L_{xy}-L_0\propto -T^2$ \cite{Ziman_2001}. The same conclusion is expected to hold for Fermi surface with hot spots since, within the conventional Boltzmann transport theory (for sufficiently weak magnetic field that can be treated perturbatively), the dominant inelastic scattering rate that governs longitudinal transport also governs transverse transport.

%the transverse Lorenz ratio serves as an equivalent probe 
%In addition, scales as  and $L_{xy}$ holds for weakly disordered FLs a
%Similarly, for FLs, the leading deviation of $L_
%Note that for FLs, the leading deviation of $L_{xy}$ agrees with that of $L$ in Eq.~\eqref{FL_WF_deviation} \cite{Ziman_2001}, and, due to the validity of the conventional Boltzmann approach, the same agreement is expected to apply for Fermi surfaces with hot spots. 

%\section{Discussion}

\section{Discussion}Naively, one may have expected the WF law to hold at $T\rightarrow 0$ only in weakly disordered Fermi liquids with well-defined quasiparticles. This is because, within the conventional Landau-Boltzmann description of transport, the universal value $L_0$ originates from integrating over Fermi functions, implying that the existence of well-defined 
%{\color{blue} nearly-free fermionic} 
quasiparticles is necessary. In contrast, as shown in this work, a broad class of weakly disordered non-Fermi liquid metals with no well-defined quasiparticles (in the sense that the electron scattering rate is either comparable to, or larger than, the energy) also satisfy the WF law at $T\rightarrow 0$. Intuitively, the fact that this class of systems obey the Wiedemann-Franz law may be understood from the fact that, while there is no well-defined Fermi surface with a sharp jump in the fermion momentum occupation function, the generalized energy distribution function $f\left(\omega\right)=-i\int\frac{d\varepsilon}{2\pi}G^{<}\left(\varepsilon,\omega\right)$, is a Fermi function (see Supplementary Material). 
%{\color{blue} As discussed in the previous section,} 
A sufficient condition for the WF law to hold is that the QBE approach is applicable; this requires, in particular, that (i) The width of electronic spectral functions at zero energy is smaller than the Fermi momentum, and that (ii) The dependence of the electronic scattering rate on momentum is non-singular. Note that, in particular, condition (i) implies that the resistivity is small compared to the Mott-Ioffe-Regel limit.

Thus, the fact the WF is obeyed at $T=0$ %{\color{blue}(as is the case in a variety of interesting ``strange metallic'' systems which show anomalous scaling laws of the resistivity vs. $T$ at $T\rightarrow 0$)} 
is not sufficient to deduce that these systems are conventional Fermi liquids in disguise. Instead, we propose to examine the deviation of the Lorenz ratio $L(T)$ from $L_0$ as $T\rightarrow 0$. Since this quantity depends on the degree of inelastic scattering, it can distinguish different sources of strange metallicity, such as Fermi liquids with a source of $T-$linear nearly-elastic scattering (such scattering from an Einstein bosonic mode whose frequency is lower than $T$), from ``true'' non-Fermi liquids where the scattering is inelastic (see Fig. \ref{fig:schematic_wfl}). 

%In this work, we proposed a criterion for strange metallicity in weakly disordered systems, using the low-temperature behavior of the Lorenz ratio.  that obey the WF law at $T\to 0$. Considering the low-$T$ leading behavior of the normalized Lorenz ratio $L/L_0 -1\propto - T^{\beta}$, we show that the exponent $\beta$ distinguishes the FL and NFL nature in a variety of different systems. We discuss the generic behavior in 2D and 3D FLs, and demonstrate the consistency of our criterion with two known mechanisms for $T$--linear resistivity. We further consider two cases of FLs with hot spots that are known to present anomalous NFL transport properties (but do not show $T$--linear resistivity at low $T$) and showed that they, too, can be diagnosed with our criterion. In addition, we constructed a solvable model of a weakly disordered MFL that obeys the WF law at low temperature. The leading deviation of this model from the WF law is $T$--linear (i.e., $\beta=1$), unlike the former cases of FLs and FLs with hot spots. {By employing the QBE approach, we obtained an additional perspective on these results, and further corroborated that Eq.~\eqref{MFL_WF_deviation} could serve as a universal criterion for strange metallicity.}  
%{\color{red} FROM HERE ***}

In practice, our criterion is applicable under experimental conditions where the electronic degrees of freedom dominate heat transport at low $T$. For the longitudinal case, while these conditions can be met in some scenarios (for example \cite{Paglione2005,smith_marginal_2008}), it could also be the case that other degrees of freedom, e.g., phonons, will dominate the thermal conductivity which would make our criterion inaccessible. To separate the electronic contribution, the transverse Lorenz ratio $L_{xy}$ is often used (since $\kappa_{xy}$ is often, although not always~\cite{grissonnanche2020chiral}, dominated by the electronic contribution). Here we showed that our criterion applies to the longitudinal and transverse cases at once, and therefore expect it to be widely applicable. 
%We leave the generalization of our criterion to the transverse Lorenz ratio to future work. 
%{\color{blue} It should be straightforward to generalize our criterion to the deviation from the transverse WF law.} We leave this to follow-up work.  

%In fact, ...

An intriguing issue concerns the %validity of the WF at $T\rightarrow 0$, as well as the behavior of the lowest-order deviation $L(T)-L_0$, 
application of our criterion to theories of quantum-critical metals, especially in cases where the electrical resistivity is $T-$linear~\cite{patel_universal_2022,wu_quantum_2022,shi_loop_2022-1}. 
%with $T-$linear resistivity 
%In particular, it would be interesting to examine these issues in theories of strange metals where the anomalous scattering rate is intrinsically related to disorder~\cite{patel_universal_2022,wu_quantum_2022} as well as in ``intrinsically clean'' critical metals in the presence of weak impurity scattering~\cite{shi_loop_2022-1}. The applicability of QBE approach in these systems is not obvious. }
%For our purposes, spatial disorder provides a physically transparent reference point.     
%Empirically, there are examples where let us point out that 
In this regard, we point out  Ref.~\cite{smith_marginal_2008}, that reported low--$T$ transport measurements in a weakly disordered 3D system at a ferrmomagnetic critical point.  
It was found that at low $T$, $\rho = \rho_0 + AT^{5/3}$ while $\rho_{\rm th} =  \rho_{\rm th,0} + BT$, such that $\overline{L} -1 \propto  - T$, consistent with MFL behavior by our criterion~\cite{Belitz2000,lohneysen_fermi-liquid_2007}. This observation is further corroborated by evidence for a $T\log\left(1/T\right)$ behavior in the specific heat~\cite{sutherland_transport_2012}, as expected for a MFL~\cite{varma_phenomenology_1989}.  

\section{Methods}

All analytical calculations are explicitly presented in the Supplementary note.

\section{Acknowledgements}We thank Sean Hartnoll, Tobias Holder, Steven Kivelson, Dmitrii Maslov, Karen Michaeli, Yuval Oreg, J\"{o}rg Schmalian, Dam T. Son, Brad Ramshaw, Sankar Das Sarma, Ady Stern, Louis Taillefer, and Senthil Todadri for useful discussions and comments on this manuscript. This work was supported by the European Research Council (ERC) under grant HQMAT (Grant Agreement No. 817799), the Israel-US Binational Science Foundation (BSF), and the Minerva Foundation. 

\section{COMPETING INTERESTS}
The authors declare no competing interests.

\section{AUTHOR CONTRIBUTIONS
}
E.T. and E.B. have contributed equally to the development of the ideas in this
work, and to the writing of the paper. E.T. did the calculations.

\section{DATA AVAILABILITY}
The data that support the findings of this study are available from the authors on
request.

\bibliographystyle{sn-standardnature}
%\bibliographystyle{natbib}

%natbib
\bibliography{refs.bib}

\appendix

\onecolumngrid

\section{Thermal conductivity in a FL near AFM criticality in 3D}
\label{HS_th_cond}
We consider the model studied in Ref.~\cite{rosch_interplay_1999}. In the framework of the variational Boltzmann approach \cite{Ziman_2001}, the low-$T$ inverse electronic thermal conductivity is given by 
\begin{eqnarray}
\frac{1}{\kappa}= \frac{1}{\kappa_{\rm imp}} + \frac{\int_{\boldsymbol{k,k'}}\left(\Phi_{\boldsymbol{k}}-\Phi_{\boldsymbol{k}'}\right)^{2}W_{\boldsymbol{kk'}}}{\left|\int_{\boldsymbol{k}}\boldsymbol{v}_{\boldsymbol{k}}\left(\varepsilon_{\boldsymbol{k}}-\mu\right)\Phi_{\boldsymbol{k}}\frac{\partial f_{\boldsymbol{k}}^{0}}{\partial\varepsilon_{\boldsymbol{k}}}\right|^{2}}.
\label{kappa_hot_spots}
\end{eqnarray}
Here $\frac{1}{\kappa_{\rm imp}}$ is the contribution from static impurities that, when combined with the residual electrical resistivity, satisfies the WF law at $T\to 0$. In addition, $\boldsymbol{v}_{\boldsymbol{k}}=\partial_{\boldsymbol{k}} \varepsilon_{\boldsymbol{k}}$ where $\varepsilon_{\boldsymbol{k}}$ is the dispersion relation, $\mu$ is the chemical potential. We are interested in the leading deviation from the WF law, associated with the second term in Eq.~\eqref{kappa_hot_spots}. We mainly follow the notation of \cite{Ziman_2001}, where the full out-of-equilibrium distribution function is given by $f_{\boldsymbol{k}} = f^0_{\boldsymbol{k}} + \Phi_{\boldsymbol{k}}\frac{\partial f^{0}_{\boldsymbol{k}}}{\partial \varepsilon_{\boldsymbol{k}}}$. $f^0$ ($n$) denotes the equilibrium Fermi (Bose) distribution function. The transition rate associated with critical AFM spin fluctuations is given by  $W_{\boldsymbol{k}\boldsymbol{k}'}=2g_{S}^{2}f_{\boldsymbol{k}}^{0}\left(1-f_{\boldsymbol{k}'}^{0}\right)n\left(\varepsilon_{\boldsymbol{k}'}-\varepsilon_{\boldsymbol{k}}\right)\text{Im}\chi \left(\boldsymbol{k}'-\boldsymbol{k},\varepsilon_{\boldsymbol{k}}-\varepsilon_{\boldsymbol{k}'}\right)$, with the retarded spin correlation function $\chi\left(\boldsymbol{q},\omega\right)=\sum_{\pm}\frac{1}{1/\left(q_{0}\xi\right)^{2}+\left(\boldsymbol{q}\pm\boldsymbol{Q}\right)^2/q_{0}^{2}-i\omega/\Gamma}$ \cite{rosch_interplay_1999}. 

To proceed we use the ansatz $\Phi_{\boldsymbol{k}}=\eta \left(\varepsilon_{\boldsymbol{k}}-\mu\right)\boldsymbol{u}\cdot\boldsymbol{k}$, with a small $\eta$ and where $\boldsymbol{u}$ is a unit-vector in the direction heat current, which is appropriate at the low-$T$ limit, where impurity scattering dominates $\kappa$. From this point the analysis is analogous to the one in Refs.~\cite{hlubina_resistivity_1995,rosch_interplay_1999}, namely, the low-$T$ scaling of $1/\kappa$ is determined by the $T$-dependent phase space of the hot-lines that scales as $\sqrt{T}$. In total, we obtain that the thermal conductivity is given by 
\begin{eqnarray}
\frac{1}{\kappa}&=& \frac{1}{\kappa_{\rm imp}} + \frac{1}{\kappa_{\rm spin}} = \frac{A}{T} + B\sqrt{T}, 
\end{eqnarray}
where $A$ and $B$ are related to the impurity and spin contributions. Finally the thermal resistivity obeys $\rho_{\rm th}= T/\kappa = \rho_{\rm th,0} + BT^{3/2}$. 

\section{Spatially extended $cf$-interaction}
\label{vertex_corrections}

We discuss the modifications to the self-energy and resistivities in the presence of a spatially extended $cf$-interaction. 
\begin{figure}[b]
\centering
\includegraphics[width=\columnwidth]{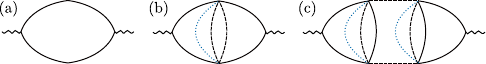}
\renewcommand{\figurename}{Supplementary Figure}
\caption{Leading diagrams in $1/N$ for the electrical and thermal current-current correlation function. Solid line correspond to $c$-fermions, Dashed lines correspond to $f$-fermions and blue dotted lines denote averaging over realizations of $V_{ijkl}$. } 
\label{fig:diagrams}
\end{figure}

\subsection{Self energy}
\label{self energy appendix}
As in the main text, we follow the notation of \cite{chowdhury_translationally_2018}. We first show that the MFL form of the $c$-fermions self-energy is unchanged by extending the range of the $cf$-interactions, up to a non-universal numerical coefficient. Let us set $\Upsilon_{\boldsymbol{r},\boldsymbol{r}'} = \delta_{\boldsymbol{r},\boldsymbol{r}'} + \eta \sum_{\boldsymbol{\boldsymbol{\delta}} = \pm \hat{x},\pm \hat{y}}  \delta_{\boldsymbol{r},\boldsymbol{r}'+\boldsymbol{\delta}}$ as in the main text. To show that the self-energy is unchanged, we note that to leading order in $\eta$, the $cf$-contribution to the self-energy of the $c$-fermions is given by  
\begin{eqnarray}
\Sigma_{cf}\left(\boldsymbol{k},i\omega\right)=-U_{cf}^{2}\int_{\boldsymbol{q},\Omega}G_{c}\left(\boldsymbol{k+q},i\omega+i\Omega\right)\widetilde{\Upsilon}\left(\boldsymbol{q}\right)\Pi_{f}\left(i\Omega\right),
\end{eqnarray}
where $\widetilde{\Upsilon}\left(\boldsymbol{q}\right)=1+4\eta\left(\cos q_{x}+\cos q_{y}\right) + \mathcal{O}\left(\eta^2\right)$. Considering $\boldsymbol{k}$ on the FS, we approximate $\varepsilon_{\boldsymbol{k}+\boldsymbol{q}}\approx v_F q \cos\theta$ where $\theta$ is the angle between $\boldsymbol{k}$ and $\boldsymbol{q}$. In addition, to obtain the most singular contribution we further approximate $\widetilde{\Upsilon}\left(\boldsymbol{q}\right)\approx1+8\eta+\mathcal{O}\left(q^{2}\right)$. Carrying the momentum and frequency integrals in the usual way (see \cite{chowdhury_translationally_2018} for details), it follows that, in the low-energy limit,
\begin{eqnarray}
\Sigma_{cf}\left(\boldsymbol{k},i\omega\right)&=-i\left(1+\alpha_{\Upsilon}\right)\frac{U_{cf}^{2}\nu_{0}}{2\pi U_{f}}\left[\log\left(\frac{U_{f}}{\left|\omega\right|}\right)-1\right]\omega
\end{eqnarray}
where $\alpha_{\Upsilon}=8\eta$, namely, the MFL form of the self-energy is unchanged. 

\subsection{Thermal current operator}

Note that the thermal current is given by $\boldsymbol{j}_{\rm th}\left(t\right)=\frac{i}{2}\sum_{l}\int_{\boldsymbol{k}}\partial_{\boldsymbol{k}}\varepsilon_{\boldsymbol{k}}\left[\dot{c}_{\boldsymbol{k}l}^{\dagger}c_{\boldsymbol{k}l}-c_{\boldsymbol{k}l}^{\dagger}\dot{c}_{\boldsymbol{k}l}\right]$ as long as the $cf$-interaction is local in the $c$-fermions. For brevity, let us show this in  a simplified 1D continuum model given by 
\begin{eqnarray}
H=\frac{1}{2m}\int_{r}\partial_{r}c_{r}^{\dagger}\partial_{r}c_{r}+\int_{r,r'}\widehat{V}_{r,r'}c_{r}^{\dagger}c_{r},
\end{eqnarray} where $\widehat{V}_{r,r'}=V\Upsilon_{r,r'}f_{r'}^{\dagger}f_{r'}$. We use the continuity equation to identify the thermal current operator: $\dot{h}_{r}=\partial_{r}j_{\text{th},r}$. Here the local Hamiltonian density is given by
\begin{eqnarray}
h_{r}=\partial_{r}c_{r}^{\dagger}\partial_{r}c_{r}+\int_{r'}\widehat{V}_{r,r'}c_{r}^{\dagger}c_{r}
\end{eqnarray}
and $\dot{h}_{r}=i\left[h_{r},H\right]$. Ignoring the time derivatives of the $f$-fermions, which do not participate in transport, we have that
\begin{eqnarray}
\dot{h}_{r}=\frac{1}{2m}\left(\partial_{r}\dot{c}_{r}^{\dagger}\partial_{r}c_{r}+\partial_{r}c_{r}^{\dagger}\partial_{r}\dot{c}_{r}\right)+\int_{r'}\widehat{V}_{r,r'}\left(\dot{c}_{rk}^{\dagger}c_{rl}+c_{rk}^{\dagger}\dot{c}_{rl}\right).
\end{eqnarray}
To proceed we rewrite the first term as $\partial_{r}\dot{c}_{r}^{\dagger}\partial_{r}c_{r}+\partial_{r}c_{r}^{\dagger}\partial_{r}\dot{c}_{r}=\partial_{r}\left(\dot{c}_{r}^{\dagger}\partial_{r}c_{r}+\partial_{r}c_{r}^{\dagger}\dot{c}_{r}\right)-\left(\dot{c}_{r}^{\dagger}\partial_{r}^{2}c_{r}+\partial_{r}^{2}c_{r}^{\dagger}\dot{c}_{r}\right)$. Then, we insert the equation of motion for $c_{r},c_{r}^{\dagger}$ (i.e. $\partial_{r}^{2}c_{r}=-i\dot{c}_{r}-\int_{r'}\widehat{V}_{r,r'}c_{r}$), which is \textit{local} as long as the $cf$-interaction is local in the $c$-fermions, and arrive at
\begin{eqnarray}
\dot{h}_{r}=\frac{1}{2m}\partial_{r}\left(\dot{c}_{r}^{\dagger}\partial_{r}c_{r}+\partial_{r}c_{r}^{\dagger}\dot{c}_{r}\right)=\partial_{r}j_{\text{th},r},
\end{eqnarray}
with $j_{\text{th},r}=\frac{1}{2m}\left(\dot{c}_{r}^{\dagger}\partial_{r}c_{r}+\partial_{r}c_{r}^{\dagger}\dot{c}_{r}\right)$. Going back to the lattice, the total thermal current is indeed given by $j_{\text{th}}=\frac{i}{2}\int_{k}\partial_{k}\varepsilon_{k}\left[\dot{c}_{k}^{\dagger}c_{k}-c_{k}^{\dagger}\dot{c}_{k}\right]$. Generalizing this to higher dimensions is straightforward. 

\subsection{Vertex corrections}
Consider the effect of the nearest-neighbors (n.n.) interaction on the conductivities. Here we demonstrate the effect on the electrical conductivity. The thermal conductivity follows exactly the same considerations. First recall that in the absence of n.n. $cf$-interaction, all vertex corrections vanish due to the locality of the $f$-fermions. The conductivities are then associated with the current-current bubble diagram shown in Fig.~\ref{fig:diagrams}a. Adding n.n. interactions lead to an additional contribution, which is associated with the bubble diagram with a single rung inserted, see Fig.~\ref{fig:diagrams}b. Importantly, while the diagram shown in Fig.~\ref{fig:diagrams}c is the same order in $1/N$, it does not contribute to the conductivities because it is non-zero only at second order in $\eta$. 

For our purposes, it is sufficient to show that the contribution of Fig.~\ref{fig:diagrams}b does not vanish. It then follows that, to leading order in $\eta$, there are vertex corrections to $\sigma$ and $\kappa$ that generically alter the high-$T$ ($T\gg \Gamma$) saturation value. The contribution due to the insertion of a single rung is given by  
\begin{eqnarray}
\delta\Pi_{J}^{x}\left(i\Omega\right)= U_{cf}^{2}\int_{\omega,\omega',\boldsymbol{k},\boldsymbol{k'}}v_{\boldsymbol{k}}^{x}v_{\boldsymbol{k'}}^{x}G_{c}\left(i\omega+i\Omega,\boldsymbol{k}\right)G_{c}\left(i\omega,\boldsymbol{k}\right)\widetilde{\Upsilon}\left(\boldsymbol{k}-\boldsymbol{k'}\right)\Pi_{f}\left(i\omega-i\omega'\right)G_{c}\left(i\omega'+i\Omega,\boldsymbol{k'}\right)G_{c}\left(i\omega',\boldsymbol{k'}\right).
\label{current_correlator}
\end{eqnarray}
We can now see that in the absence of n.n. interaction, i.e., when $\widetilde{\Upsilon}=1$, this contribution vanishes since the integrand is odd in $\boldsymbol{k}$ and in $\boldsymbol{k}'$. However upon introducing the n.n. interaction, there is an additional odd part in $\boldsymbol{k}$ and in $\boldsymbol{k}'$:
\begin{eqnarray}
\widetilde{\Upsilon}\left(\boldsymbol{k}-\boldsymbol{k'}\right)&=4\eta\sin k_{x}\sin k'_{x}+\left(\text{terms even in \ensuremath{k_{x},k'_{x}}}\right).
\end{eqnarray}
The product of the odd part of $\widetilde{\Upsilon}$ with the velocities is even in $\boldsymbol{k}$ and $\boldsymbol{k}'$ and therefore does not vanish under the integral. Note that only taking the leading diagrams in $\eta$ is equivalent to a perturbative treatment of the current vertex, which justified since we are perturbing about a non-singular point, i.e., since the conductivities are finite at $\eta=0$.

\subsection{High-$T$ saturation value}

We recall the Kubo formula for the dc electrical conductivity, expressed in terms of the analytic continuation of the Matsubara frequency current-current correlator:
\begin{eqnarray}
\sigma\left(T\right)=\lim_{\Omega\to0}\lim_{\eta\to0}\frac{\text{Im}\Pi_{J}^{x}\left(i\Omega_{l}\to\Omega+i\eta\right)}{\Omega}.
\end{eqnarray}
For any $T>0$, we can thus define dimensionless frequencies $\overline{\Omega}_{l}=\Omega_{l}/T$
and $\overline{\Omega}=\Omega/T$ such that 
\begin{eqnarray}
\sigma\left(T\right)=\lim_{\overline{\Omega}\to0}\lim_{\eta\to0}\frac{\text{Im}\Pi_{J}^{x}\left(i\overline{\Omega}_{l}\to\overline{\Omega}+i\eta\right)}{T\overline{\Omega}}.
\label{rescaled_kubo}
\end{eqnarray}
Therefore, obtaining the $T$-scaling of $\text{Im}\Pi_{J}^{x}\left(i\overline{\Omega}_{l}\to\overline{\Omega}+i\eta\right)$ yields the $T$-scaling of $\sigma$. We will now show that this simple power counting argument can be used to obtain the $T$-scaling of the conductivity correction. 

To begin, consider the expression for the current correlator related to the conductivity correction:
\begin{eqnarray}
\delta\Pi_{J}^{x}\left(i\Omega_{l}\right)&=&U_{cf}^{2}T^{2}\sum_{m,n\in\mathbb{Z}}\int_{\boldsymbol{k},\boldsymbol{k'}}v_{\boldsymbol{k}}^{x}v_{\boldsymbol{k'}}^{x}\\
&&G_{c}\left(i\omega_{n}+i\Omega_{l},\boldsymbol{k}\right)G_{c}\left(i\omega_{n},\boldsymbol{k}\right)\widetilde{\Upsilon}\left(\boldsymbol{k}-\boldsymbol{k}'\right)\Pi_{f}\left(i\omega_{n}-i\omega_{m}\right)G_{c}\left(i\omega_{m}+i\Omega_{l},\boldsymbol{k'}\right)G_{c}\left(i\omega_{m},\boldsymbol{k'}\right)
\end{eqnarray}
We take the odd part of $\widetilde{\Upsilon}$, as above, and approximate
$\sin k_{x}\sim\frac{1}{W_{c}}v_{\boldsymbol{k}}^{x}$:
\begin{eqnarray}
\delta\Pi_{J}^{x}\left(i\Omega\right)&\sim&\eta\frac{U_{cf}^{2}}{W_{c}^{2}}T^{2}\sum_{m,n\in\mathbb{Z}}\int_{\boldsymbol{k},\boldsymbol{k'}}\left(v_{\boldsymbol{k}}^{x}\right)^{2}\left(v_{\boldsymbol{k'}}^{x}\right)^{2}\\
&&G_{c}\left(i\omega_{n}+i\Omega_{l},\boldsymbol{k}\right)G_{c}\left(i\omega_{n},\boldsymbol{k}\right)\Pi_{f}\left(i\omega_{n}-i\omega_{m}\right)G_{c}\left(i\omega_{m}+i\Omega_{l},\boldsymbol{k'}\right)G_{c}\left(i\omega_{m},\boldsymbol{k'}\right).
\end{eqnarray}
This allows us to change the momentum integration to energy integration in the usual way. Then, by expressing the Green's functions with the spectral representation
and integrating over $\varepsilon,\varepsilon'$, we obtain
\begin{eqnarray}
\delta\Pi_{J}^{x}\left(i\Omega_{l}\right) & \sim&\eta\frac{U_{cf}^{2}}{W_{c}^{2}}\nu_{0}^{2}v_{F}^{4}\int_{\epsilon_{1},\epsilon_{2},\epsilon_{3},\epsilon_{3},\epsilon_{5}}\mathcal{S}_{c}\left(\epsilon_{1},\epsilon_{2}\right)\mathcal{F}\left(\epsilon_{3}\right)\mathcal{S}_{c}\left(\epsilon_{4},\epsilon_{5}\right).\\
 && \times T^{2}\sum_{n,m\in\mathbb{Z}}\frac{1}{\epsilon_{1}-i\Omega_{l}-i\omega_{n}}\frac{1}{\epsilon_{2}-i\omega_{n}}\frac{1}{\epsilon_{3}-i\omega_{n}+i\omega_{m}}\frac{1}{\epsilon_{4}-i\Omega_{l}-i\omega_{m}}\frac{1}{\epsilon_{5}-i\omega_{m}},
\end{eqnarray}
where 
\begin{eqnarray}
\mathcal{S}_{c}\left(\epsilon_{i},\epsilon_{j}\right)\equiv \frac{{\Sigma_{R}^{''}\left(\epsilon_{i}\right)}+\Sigma_{R}^{''}\left(\epsilon_{j}\right)}{\left(\epsilon_{i}-\epsilon_{j}+{\Sigma_{R}^{'}\left(\epsilon_{i}\right)}-\Sigma_{R}^{'}\left(\epsilon_{j}\right)\right)^{2}+\left({\Sigma_{R}^{''}\left(\epsilon_{i}\right)}+\Sigma_{R}^{''}\left(\epsilon_{j}\right)\right)^{2}}
\end{eqnarray}
and $\mathcal{F}={\rm Im}\Pi_{f}$ is obtained from a convolution of two real-time Green's
functions of the $f$-fermions. Using the scaling (SYK) form of the
$f$-fermions, one can write
\begin{eqnarray}
\mathcal{F}\left(\epsilon\right)=\frac{1}{U_{f}}\overline{\mathcal{F}}\left(\overline{\epsilon}\right),
\label{T_scaling_of_Pi_f}
\end{eqnarray}
where $\overline{\mathcal{F}}$ is a dimensionless function of the
dimensionless variable $\overline{\epsilon}=\frac{\epsilon}{T}$.
Then, by rescaling $\epsilon_{1},...,\epsilon_{5},\omega_{n},\omega_{m}$
and $\Omega_{l}$ by $T$, we see that 
\begin{eqnarray}
\delta\Pi_{J}^{x}\left(i\overline{\Omega}_{l}\right) & \sim&\eta\frac{U_{cf}^{2}}{U_{f}W_{c}^{2}}\nu_{0}^{2}v_{F}^{4}\int_{\overline{\epsilon}_{1},\overline{\epsilon}_{2},\overline{\epsilon}_{3},\overline{\epsilon}_{4},\overline{\epsilon}_{5}}\mathcal{S}_{c}\left(\epsilon_{1},\epsilon_{2}\right)\overline{\mathcal{F}}\left(\overline{\epsilon}_{3}\right)\mathcal{S}_{c}\left(\epsilon_{4},\epsilon_{5}\right).\\
 && \times T^{2}\sum_{n,m\in\mathbb{Z}}\frac{1}{\overline{\epsilon}_{1}-i\overline{\Omega}_{l}-i\overline{\omega}_{n}}\frac{1}{\overline{\epsilon}_{2}-i\overline{\omega}_{n}}\frac{1}{\overline{\epsilon}_{3}-i\overline{\omega}_{n}+i\overline{\omega}_{m}}\frac{1}{\overline{\epsilon}_{4}-i\overline{\Omega}_{l}-i\overline{\omega}_{m}}\frac{1}{\overline{\epsilon}_{5}-i\overline{\omega}_{m}}.
\end{eqnarray}
Here the overline denotes the rescaled variables.

To obtain the scaling behavior of $\delta\sigma$ in the limit $T\gg\Gamma$,
we set $\Gamma=0$, such that the self-energy has the scaling MFL
form, $\Sigma_{R}\left(\epsilon_{i}\right)\approx T\overline{\Sigma}_{cf,R}\left(\overline{\epsilon}_{i}\right)$, which means that 
\begin{eqnarray}
\mathcal{S}_{c}\left(\epsilon_{i},\epsilon_{j}\right)=\frac{1}{T}\overline{\mathcal{S}}_{c}\left(\overline{\epsilon}_{i},\overline{\epsilon}_{j}\right).
\end{eqnarray}
As before, the overline denotes dimensionless functions and variables. Inserting
this back into the current correlator, we obtain that
\begin{eqnarray}
\delta\Pi_{J}^{x}\left(i\overline{\Omega}_{l}\right) & \sim&\eta\frac{U_{cf}^{2}}{U_{f}W_{c}^{2}g^{4}}\nu_{0}^{2}v_{F}^{4}\int_{\overline{\epsilon}_{1},\overline{\epsilon}_{2},\overline{\epsilon}_{3},\overline{\epsilon}_{4},\overline{\epsilon}_{5}}\overline{\mathcal{S}}_{c}\left(\overline{\epsilon}_{1},\overline{\epsilon}_{2}\right)\overline{\mathcal{F}}\left(\overline{\epsilon}_{3}\right)\overline{\mathcal{S}}_{c}\left(\overline{\epsilon}_{4},\overline{\epsilon}_{5}\right).\\
 && \times\sum_{n,m\in\mathbb{Z}}\frac{1}{\overline{\epsilon}_{1}-i\overline{\Omega}_{l}-i\overline{\omega}_{n}}\frac{1}{\overline{\epsilon}_{2}-i\overline{\omega}_{n}}\frac{1}{\overline{\epsilon}_{3}-i\overline{\omega}_{n}+i\overline{\omega}_{m}}\frac{1}{\overline{\epsilon}_{4}-i\overline{\Omega}_{l}-i\overline{\omega}_{m}}\frac{1}{\overline{\epsilon}_{5}-i\overline{\omega}_{m}}.
\end{eqnarray}
We can now observe that $\delta\Pi_{J}^{x}\left(i\overline{\Omega}_{l}\right)$
is independent of $T$. Using Eq.~\eqref{rescaled_kubo}, it follows that 
\begin{eqnarray}
\delta\sigma\sim\eta\frac{U_{f}v_{F}^{4}}{W_{c}^{2}U_{cf}^{2}\nu_{0}^{2}}\times\frac{1}{T}\times\left(\text{dimensionless integral}\right).
\end{eqnarray}
Hence the correction $\delta\sigma$ scales as $1/T$, similarly to $\sigma$. For the thermal conductivity, the derivation is similar, and gives $\delta\kappa\sim {\rm const}$ when $T\gg \Gamma$. The saturation value of the Lorenz ratio is thus
altered by vertex corrections.

\section{Quantum Boltzmann equation}
\label{QBE_app}

Here we outline the idea behind the QBE for the generalized Fermi distribution function, following similar steps to Refs.~\cite{nave_transport_2007,mahan_many-particle_2000}. 
\subsection{Derivation of QBE}
We begin by introducing the non-equilibrium Green's functions $\tilde{G}$ and $\tilde{\Sigma}$ that satisfy the Dyson's equation:
\begin{eqnarray}
\tilde{G}=\tilde{G}_{0}+\tilde{G}_{0}\tilde{\Sigma}\tilde{G}
\label{Dyson_eq}
\end{eqnarray}
where
\begin{eqnarray}
\tilde{G}=\begin{pmatrix}G_{t} & -G^{<}\\
G^{>} & -G_{\bar{t}}
\end{pmatrix}
\end{eqnarray}
and similarly for $\tilde{\Sigma}$. $G_0$ denotes the non-interacting Green's function. As in \cite{mahan_many-particle_2000}, the multiplication in Eq.~\eqref{Dyson_eq} denotes integration over a shared space-time variable. In addition, 
\begin{eqnarray}
G^{>}\left(1,2\right)&=&-i\left\langle c\left(1\right)c^{\dagger}\left(2\right)\right\rangle; \\G^{<}\left(1,2\right)&=&i\left\langle c^{\dagger}\left(2\right)c\left(1\right)\right\rangle; \\G_{t}\left(1,2\right)&=&\Theta\left(t_{1}-t_{2}\right)G^{>}\left(1,2\right)+\Theta\left(t_{2}-t_{1}\right)G^{<}\left(1,2\right);\\G_{\bar{t}}\left(1,2\right)&=&\Theta\left(t_{2}-t_{1}\right)G^{>}\left(1,2\right)+\Theta\left(t_{1}-t_{2}\right)G^{<}\left(1,2\right),
\end{eqnarray}
where $1=\left(\boldsymbol{r}_{1},t_{1}\right)$ and similarly for 2. Note that 
\begin{eqnarray}
G^{R}&=G_{t}-G^{<}=G^{>}-G_{\bar{t}};\\G^{A}&=G_{t}-G^{>}=G^{<}-G_{\bar{t}},
\end{eqnarray}
where $G^{R(A)}$ is the familiar retarded (advanced) Green's functions. The derivation proceeds by changing to center-of-mass and relative coordinates: %{\color{red} $T$ can be confused for the temperature? Replace with $\overline{t}$?}
\begin{eqnarray}
\left(\boldsymbol{R},\overline{t}\right)=\frac{1}{2}\left(1+2\right),\quad\left(\boldsymbol{r},t\right)=1-2.
\end{eqnarray}
To derive a self-consistent Dyson's equation for $G^<$, we consider the Fourier transform with respect to the relative coordinates:
\begin{eqnarray}
\tilde{G}\left(\boldsymbol{k},\omega,\boldsymbol{R},\overline{t}\right)=\int_{\boldsymbol{r}}e^{i\boldsymbol{k}\boldsymbol{r}}\int_{t}e^{i\omega t}\tilde{G}\left(\boldsymbol{r},t,\boldsymbol{R},\overline{t}\right).
\end{eqnarray}
In the following we slightly abuse the notation above by relabelling the center-of-mass coordinates by $\boldsymbol{r}$ and $t$. In addition, we will occasionally ignore these arguments for brevity. Note that in thermal equilibrium, 
\begin{eqnarray}
G^{<}\left(\boldsymbol{k},\omega\right)&=&if_{0}\left(\omega\right)\mathcal{A}\left(\boldsymbol{k},\omega\right)\\G^{>}\left(\boldsymbol{k},\omega\right)&=&-i\left(1-f_{0}\left(\omega\right)\right)\mathcal{A}\left(\boldsymbol{k},\omega\right)
\end{eqnarray}
where $f_0$ is the equilibrium Fermi distribution at some temperature, and the spectral function $\mathcal{A}\left(\boldsymbol{k},\omega\right)=-i\left(G^{R}\left(\boldsymbol{k},\omega\right)-G^{A}\left(\boldsymbol{k},\omega\right)\right)$. 

Following \cite{nave_transport_2007,mahan_many-particle_2000}, the Dyson equation for $G^<$ is given by 
\begin{eqnarray}
\left[\omega-\epsilon_{\boldsymbol{k}}-\text{Re}\Sigma^{R},G^{<}\right]-\left[\Sigma^{<},\text{Re}G^{R}\right]=\Sigma^{>}G^{<}-G^{>}\Sigma^{<},
\label{equation_for_G_lesser}
\end{eqnarray}
where the multiplication here is standard, e.g., $\Sigma^{>}G^{<}=\Sigma^{>}\left(\boldsymbol{k},\omega,\boldsymbol{r},t\right)G^{<}\left(\boldsymbol{k},\omega,\boldsymbol{r},t\right)$, and we have introduced the generalized Poisson brackets, 
\begin{eqnarray}
\left[A,B\right]=\partial_{\omega}A\partial_{t}B-\partial_{t}A\partial_{\omega}B+\nabla_{\boldsymbol{r}}A\cdot\nabla_{\boldsymbol{k}}B-\nabla_{\boldsymbol{k}}A\cdot\nabla_{\boldsymbol{r}}B.
\end{eqnarray}

To derive the QBE, we use the fact that $\Sigma_R$ is momentum independent in the low-energy limit and note that the spectral function of $c$-fermions, 
\begin{equation}
    \mathcal{A}_{c}\left(\omega,\boldsymbol{k}\right)=\frac{-2\text{Im}\Sigma_{R}\left(\omega\right)}{\left(\omega-\varepsilon_{\boldsymbol{k}}-\text{Re}\Sigma_{R}\left(\omega\right)\right)^{2}+\left(\text{Im}\Sigma_{R}\left(\omega\right)\right)^{2}},
\end{equation}
is sharply peaked as a function of $\varepsilon_{\boldsymbol{k}}$ at the FS for sufficiently small $\omega$. In addition, we note that $\int \frac{d\varepsilon}{2\pi}\mathcal{A}_c = 1$. Assuming that these features persist if the system is sufficiently close to local equilibrium (limiting ourselves to linear response), it follows that by changing the integration variables from $\boldsymbol{k}$ to $\varepsilon\equiv \varepsilon_{\boldsymbol{k}}$ and $\hat{\boldsymbol{k}}$ we can define a generalized distribution function,
\begin{eqnarray}
f_{c}\left(\hat{\boldsymbol{k}},\omega,\boldsymbol{r},t\right)\equiv-i\int\frac{d\varepsilon}{2\pi}G_{c}^{<}\left(\hat{\boldsymbol{k}},\varepsilon,\omega,\boldsymbol{r},t\right), \label{gen_dist_function}
\end{eqnarray}
that describes the distribution of $c$-fermions with energy $\omega$ at position $\boldsymbol{r}$ and time $t$. Similarly, $1-f_c\left(\hat{\boldsymbol{k}},\omega,\boldsymbol{r},t\right)=i\int\frac{d\varepsilon}{2\pi}G_c^{>}\left(\varepsilon,\omega,\boldsymbol{r},t\right)$. 

By defining the generalized distribution function \eqref{gen_dist_function} together with the above assumptions we effectively restrict the momentum of the $c$-fermions to the FS, an approximation known as the Prange-Kadanoff (PK) reduction scheme. %Roughly speaking, it is consistent as long as the available scattering mechanisms are sufficiently smooth in momentum space, such that scattering events in the normal direction to the FS are suppressed due to their small spectral weight. 
The PK reduction can be consistently applied provided that the spectral function is sharply peaked at the FS as a function of $\varepsilon$. More precisely, assuming that the typical momentum transfer due to different scattering mechanisms can be characterized by a ball of radius $q_*(T)$, the PK reduction is consistent if $|\text{Im}\Sigma_{R}\left(\omega \lesssim T\right) | \ll v_F q_*(T)$ for all scattering mechanisms. 
%; and (ii) the momentum dependence of the single-particle scattering rate sufficiently weak. 
In our case, the consistency of the PK reduction at low $T$ follows from the locality of the $f$-fermions, namely, $q_* = k_F$ and indeed in the $T$-window of interest we have that $\Gamma,T\ll v_F k_F \sim E_F$. More generally, we can see that in the presence of disorder scattering, the PK reduction is consistent if the momentum dependence of the single-particle scattering rate is sufficiently weak.
%\footnote{Note that weakly disordered MFLs in certain Yukawa-type models where scattering off critical bosons is sharply peaked at $\boldsymbol{q}\sim 0$, as studied in Ref.~\cite{guo_large_2022}, serves as an example where this condition is not met.}.

%In addition, (ii) is satisfied since scattering off of the local $f$-fermions is completely momentum-independent. 

To obtain the QBE for $f_c$, we consider the equation of motion for $G^<$. Restricting ourselves to slowly varying force fields, the QBE is obtained from a gradient expansion~\cite{mahan_many-particle_2000} followed by an integration over $\varepsilon$ of Eq.~\eqref{equation_for_G_lesser} \cite{nave_transport_2007}, from which we arrive at
\begin{eqnarray}
\mathcal{D}f_{c}\left(\hat{\boldsymbol{k}},\omega,\boldsymbol{r},t\right)=\mathcal{I}_{{\rm coll}}\left[f_{c}\left(\hat{\boldsymbol{k}},\omega,\boldsymbol{r},t\right) \right],
\end{eqnarray}
where we have introduced the differential operator
\begin{eqnarray}
\mathcal{D}\equiv\left(1-\partial_{\omega}{\rm Re}\Sigma_{R}\right)\partial_{t}+\partial_{t}{\rm Re}\Sigma_{R}\partial_{\omega}\nonumber-\nabla_{\boldsymbol{r}}{\rm Re}\Sigma_{R}\cdot\nabla_{\boldsymbol{k}_{F}}+\nabla_{\boldsymbol{k}_{F}}\left(\varepsilon_{\boldsymbol{k}}+{\rm Re}\Sigma_{R}\right)\cdot\nabla_{\boldsymbol{r}},
\end{eqnarray}
and $\nabla_{{\boldsymbol{k}_F}}g \equiv \nabla_{{\boldsymbol{k}}}g|_{\boldsymbol{k}\in{\rm FS}}$ for some function $g$.
The collision integral is given by $\mathcal{I}_{\rm coll} = \int_{\varepsilon_{\boldsymbol{k}}} \Sigma^{>}G^{<}-G^{>}\Sigma^{<}$. Note that $\mathcal{I}_{\rm coll} = \mathcal{I}_{\rm dis} + \mathcal{I}_{cf}$, where $\mathcal{I}_{\rm dis}$ and $\mathcal{I}_{cf}$ denotes the contributions due to disorder and the $cf$-interactions, respectively. Explicitly, the self-energies are given by 
\begin{eqnarray}
\Sigma_{\text{dis}}^{<\left(>\right)}\left(\boldsymbol{k},\omega,\boldsymbol{r},t\right)&=\int_{\boldsymbol{q}}W_{\boldsymbol{k}-\boldsymbol{q}}^{2}G^{<\left(>\right)}\left(\boldsymbol{q},\omega,\boldsymbol{r},t\right),
\end{eqnarray}
and
\begin{eqnarray}
\Sigma_{cf}^{<}\left(\boldsymbol{k},\omega,\boldsymbol{r},t\right)&=&U_{cf}^{2}\int_{\boldsymbol{q},\nu}\widetilde{\Upsilon}\left(\boldsymbol{q}\right)\text{Im}\Pi_{f}^{R}\left(\nu\right)\left\{ \left(n_{0}\left(\nu\right)+1\right)G^{<}\left(\boldsymbol{k}+\boldsymbol{q},\omega+\nu\right)+n_{0}\left(\nu\right)G^{<}\left(\boldsymbol{k}+\boldsymbol{q},\omega-\nu\right)\right\} \\
\Sigma_{cf}^{>}\left(\boldsymbol{k},\omega,\boldsymbol{r},t\right)&=&U_{cf}^{2}\int_{\boldsymbol{q},\nu}\widetilde{\Upsilon}\left(\boldsymbol{q}\right)\text{Im}\Pi_{f}^{R}\left(\nu\right)\left\{ n_{0}\left(\nu\right)G^{>}\left(\boldsymbol{k}+\boldsymbol{q},\omega+\nu\right)+\left(n_{0}\left(\nu\right)+1\right)G^{>}\left(\boldsymbol{k}+\boldsymbol{q},\omega-\nu\right)\right\} 
\end{eqnarray}
Here, $n_0$ is the equilibrium Bose distribution function. We allow spatially correlated disorder. In addition, we allow for a spatially extended $cf$-interaction which introduces a momentum dependence to the scattering amplitude from the $f$-fermions, $\widetilde{\Upsilon}\left(\boldsymbol{q}\right)$ (as in the Vertex correction section above). We discuss the validity of the PK reduction in more general terms below.
%To consistently apply the PK reduction, we restrict ourselves to sufficiently smooth functions $\widetilde{\Upsilon}\left(\boldsymbol{q}\right)$, i.e., sufficiently short-range interactions. 
For our current discussion, $\widetilde{\Upsilon} = 1$ and $W^2_{\boldsymbol{k}-\boldsymbol{q}} = W^2$ which is independent of momentum. These general forms of the different parts of the self-energies will be useful later on. The collision integrals are obtained from the above via an integration over $\varepsilon_{\boldsymbol{k}}$, for example, %{\color{red}Mention here already the crucial step of integrating over $\varepsilon_{\bm{k}}$ in the expression for $\mathcal{I}_{coll}$. How do we see that the scattering rate $\Pi_{cf}$ needs to be smooth in $q$? It seems that $\varepsilon_{\bm{k}}$ depends only on the self-energy, so having a smooth self-energy is sufficient.} {\color{blue} The integral over $\varepsilon_k$ can be carried out consistently only if $\Upsilon$ is sufficiently smooth because it determines the characteristic momentum transfer. For example if $\Upsilon = \delta(q)$ the PK reduction is not valid, nor is MFL form of the self energy. I think the two statements are equivalent in our case.} 
\begin{eqnarray}
\label{collision integral}
\mathcal{I}_{cf}&=&\nu_{0}U_{cf}^{2}\int_{\hat{\boldsymbol{k}}',\omega',\boldsymbol{q},\nu}\tilde{\Upsilon}\left(\boldsymbol{q}\right)\text{Im}\Pi_{f}^{R}\left(\nu\right)\times\delta\left(k_{F}\hat{\boldsymbol{k}'}-k_{F}\hat{\boldsymbol{k}}-\boldsymbol{q}\right)\\&\times&\bigg(\delta\left(\omega'-\omega-\nu\right)\bigg\{ n_{0}\left(\nu\right)\left[1-f\left(\hat{\boldsymbol{k}'},\omega'\right)\right]f\left(\hat{\boldsymbol{k}},\omega\right)-\left[1+n_{0}\left(\nu\right)\right]f\left(\hat{\boldsymbol{k}'},\omega'\right)\left[1-f\left(\hat{\boldsymbol{k}},\omega\right)\right]\bigg\}\nonumber \\&+&\delta\left(\omega'-\omega+\nu\right)\bigg\{\left[1+n_{0}\left(\nu\right)\right]\left[1-f\left(\hat{\boldsymbol{k}'},\omega'\right)\right]f\left(\hat{\boldsymbol{k}},\omega\right)-n_{0}\left(\nu\right)f\left(\hat{\boldsymbol{k}'},\omega'\right)\left[1-f\left(\hat{\boldsymbol{k}},\omega\right)\right]\bigg\}\bigg).\nonumber
\end{eqnarray}
At this point we can explicitly see that the momentum of the $c$-fermions is restricted to the Fermi surface: $\mathcal{I}_{cf}= \mathcal{I}_{cf}\left(\hat{\boldsymbol{k}},\omega,\boldsymbol{r},t \right)$ (and similarly for $\mathcal{I}_{\rm dis}$). 

It is worthwhile to comment on two subtle points in the above derivation: (i) The elimination of the second term on the LHS of Eq.~\eqref{equation_for_G_lesser} (i.e. $\left[\Sigma^{<},\text{Re}G^{R}\right]$) is due to the fact that we have assumed a particle-hole symmetric form for the density of states 
(i.e., a constant DOS $\nu(\varepsilon) \approx \nu_0$, extending from $-W_c/2$ to $W_c/2$, where $W_c$ is the itinerant electron bandwidth). For a more generic DOS, the corresponding correction to this approximation is of the order of $|\Sigma^<|/W_c$. The QBE is valid when this correction is small, namely, when the scattering rate is small compared to the Fermi energy, similarly to the standard Boltzmann equation. (ii) Note that in order to integrate over $\varepsilon_{\boldsymbol{k}}$ in the collision integral, we have used the fact that the scattering rate depends weakly on $\varepsilon_{\boldsymbol{k}}$ at low energies. This is consistent with the fact that the internal frequency, $\nu$, is always restricted to be of the order of $T$, given that the external frequency $\omega \sim T$ and the system is close to thermal equilibrium. 
 
%{\color{red}Discuss subtleties: 1. the elimination of the second term in the RHS of (51) is due to the fact that we have assumed a particle-hole symmetric form for the DOS. If we haven't done that, we would get a corretion of the order of $\Sigma^</W$, where $W$ is the bandwidth. The Boltzmann equation is valid when this is small, i.e., the scattering rate is small compared to the Fermi energy. 2. In the integration over $\varepsilon$ in the collision integral, we have used the fact that the scattering rate depends weakly on $\varepsilon$. In addition, we not that the internal frequency $\nu$ is always restricted to be of order $T$, given that $\omega\sim T$ and that we are close to equilibrium.}

\subsection{Variational formulation}

A direct solution of this QBE is clearly a non-trivial task. Instead, we will compute the resitivities via a variational formulation of the QBE. The validity of the WF law at $T\to0$ essentially follows from the fact that the
elastic scattering term, $\mathcal{I}_{\rm dis}$, dominates the inelastic term $\mathcal{I}_{cf}$, similarly to the conventional QBE description of FLs. To demonstrate this, we linearize the QBE in a manner that allows us to utilize a variational formulation of the QBE along the lines of \cite{Ziman_2001,nave_transport_2007}. We parameterize the deviation from equilibrium with the function $\phi\left(\hat{\boldsymbol{k}},\omega,\boldsymbol{r},t\right)$, such that the full distribution function is approximated as follows,
\begin{eqnarray}
f_c\left(\hat{\boldsymbol{k}},\omega,\boldsymbol{r},t\right) = f_0\left(\omega,\boldsymbol{r},t\right) - \phi\left(\hat{\boldsymbol{k}},\omega,\boldsymbol{r},t\right) \partial_{\omega}f_0\left(\omega,\boldsymbol{r},t\right).
\label{param_of_fermi_fcn}
\end{eqnarray}
The local equilibrium distribution $f_0$ nullifies the collision integrals by definition. Also note that $f_0$ could depend on space and time via local temperature $1/\beta\left(\boldsymbol{r},t\right)$ or chemical potential $\mu\left(\boldsymbol{r},t\right)$ \cite{nave_transport_2007}. Using the form \eqref{param_of_fermi_fcn}, and following the steps in Ref.~\cite{nave_transport_2007}, one can show that the linearized QBE in the presence of a uniform electric field $\boldsymbol{E}$, which we will consider for the computation of the electrical resistivity, is given by 
\begin{eqnarray}
-\boldsymbol{E}\cdot \hat{\boldsymbol{k}}v_F \partial_{\omega}f_0 = \mathcal{I_{\rm coll}}.
\end{eqnarray}
And similarly for an applied uniform thermal gradient, the linearized QBE we will consider for the computation of the thermal resistivity is given by 
\begin{eqnarray}
\nabla_{\boldsymbol{k}_F}\varepsilon_{\boldsymbol{k}}\cdot \nabla_{\boldsymbol{r}}f_0 = \mathcal{I_{\rm coll}}.
\end{eqnarray}

Remarkably, the form \eqref{param_of_fermi_fcn} enables us to relate the thermal and electrical resistivities to a variational problem in the function $\phi$. Specifically, the physical values of $\rho$ and $\rho_{\rm th}$ correspond to $\phi$ that minimizes the functionals ($a={\rm el,th}$) $\mathcal{F}^{a}[\phi] = \mathcal{F}_{\rm dis}^{a}[\phi] + \mathcal{F}_{cf}^{a}[\phi]$, where the disorder and $cf$-interaction contributions to the resistivities are denoted by $\mathcal{F}_{\rm dis}$ and $\mathcal{F}_{cf}$, respectively. The derivation of the variational formulation is analogous to the one in Refs.~\cite{nave_transport_2007,Ziman_2001}. We therefore state the final expressions for the resistivities. To do so, we must introduce an inner product defined as
\begin{eqnarray}
\left\langle g,h\right\rangle \equiv\nu_{0}\int_{\hat{\boldsymbol{k}},\omega}g\left(\hat{\boldsymbol{k}},\omega\right)h\left(\hat{\boldsymbol{k}},\omega\right)
\end{eqnarray}
for some functions $g$ and $h$. In addition we define the operators $\mathcal{P}_a$ as
\begin{eqnarray}
\mathcal{P}_a\phi\equiv\nu_{0}\int_{\hat{\boldsymbol{k}}',\omega'}\left(\phi\left(\hat{\boldsymbol{k}},\omega\right)-\phi\left(\hat{\boldsymbol{k}}',\omega'\right)\right)P_a\left(\hat{\boldsymbol{k}},\omega,\hat{\boldsymbol{k}}',\omega'\right)
\end{eqnarray}
with $P_a$ being the equilibrium transitional rates related to the different scattering mechanisms. Using these definitions, the electrical resistivities are conveniently given by 
\begin{eqnarray}
\mathcal{F}^{a}\left[\phi\right]=\mathcal{F}_{\text{dis}}^{a}\left[\phi\right]+\mathcal{F}_{cf}^{a}\left[\phi\right]=\frac{\left\langle \phi,\mathcal{P}_{\text{dis}}\phi\right\rangle }{\left|\left\langle \phi,X_{\text{ }}^{a}\right\rangle \right|^{2}}+\frac{\left\langle \phi,\mathcal{P}_{cf}\phi\right\rangle }{\left|\left\langle \phi,X_{\text{ }}^{a}\right\rangle\right| ^{2}}, \quad a={\rm el,th}.
\label{var_formulation}
\end{eqnarray}
Here, 
\begin{eqnarray}
\left\langle \phi,\mathcal{P}_{\text{dis}}\phi\right\rangle &=&\nu_{0}\beta\int_{\omega,\hat{\boldsymbol{k}},\omega',\hat{\boldsymbol{k}'}}W^{2}\left(\phi\left(\hat{\boldsymbol{k}},\omega\right)-\phi\left(\hat{\boldsymbol{k}'},\omega'\right)\right)^{2}\delta\left(\omega-\omega'\right)f_{0}\left(\omega\right)\left(1-f_{0}\left(\omega'\right)\right);
\end{eqnarray}
\begin{eqnarray}
\label{factor of 2}
    \left\langle \phi,\mathcal{P}_{cf}\phi\right\rangle &=&2U_{cf}^2\nu_{0}\beta\int_{\omega,\hat{\boldsymbol{k}},\omega',\hat{\boldsymbol{k}'},\nu,\boldsymbol{q}}\tilde{\Upsilon}\left(\boldsymbol{q}\right)\text{Im}\Pi_{f}^{R}\left(\nu\right)f_{0}\left(\omega\right)\left(1-f_{0}\left(\omega'\right)\right)n_{0}\left(\nu\right) \\
&\times&\left(\phi\left(\hat{\boldsymbol{k}},\omega\right)-\phi\left(\hat{\boldsymbol{k}'},\omega'\right)\right)^{2}\delta\left(k_{F}\hat{\boldsymbol{k}'}-k_{F}\hat{\boldsymbol{k}}-\boldsymbol{q}\right)\delta\left(\omega'-\omega-\nu\right), \nonumber
\end{eqnarray}
where the factor of 2 comes from the equal contribution of an emission and absorption of a `$\Pi_f$-boson' \cite{nave_transport_2007}, and
\begin{eqnarray}
\left|\left\langle \phi,X^{\text{el}}\right\rangle\right|^2 &=&\left|\nu_{0}\int_{\hat{\boldsymbol{k}},\omega}v_{F}\hat{\boldsymbol{k}}\phi\left(\hat{\boldsymbol{k}},\omega\right)\partial_{\omega}f_{0}\left(\omega\right)\right|^2; \\
\left|\left\langle \phi,X^{\text{th}}\right\rangle\right|^2 &=&\left|\nu_{0}\int_{\hat{\boldsymbol{k}},\omega}v_{F}\hat{\boldsymbol{k}}\phi\left(\hat{\boldsymbol{k}},\omega\right)\omega\partial_{\omega}f_{0}\left(\omega\right)\right|^2.
\label{phiX}
\end{eqnarray}
 
Anticipating the dominance of elastic scattering as $T\to 0$, we may use the variational ansatzes:
\begin{eqnarray}
\phi^{\text{el}}_{\rm dis}\left(\hat{\boldsymbol{k}},\omega\right)=\eta\boldsymbol{u}\cdot\hat{\boldsymbol{k}},\quad\phi^{\text{th}}_{\rm dis}\left(\hat{\boldsymbol{k}},\omega\right)=\eta\omega\boldsymbol{v}\cdot\hat{\boldsymbol{k}},
\label{ansatzes}
\end{eqnarray}
where $\boldsymbol{u}$ ($\boldsymbol{v}$) denotes a unit vector in the direction of the electrical (heat) current and $\eta$ is a small parameter. By inserting the ansatzes \eqref{ansatzes} into Eq.~\eqref{var_formulation} with the above definitions, we can explicitly see that the validity of the WF law at $T \to 0$ follows from the dominance of the $T$-independent elastic scattering contribution, $\mathcal{F}_{\rm dis}$, over the inelastic term, $\mathcal{F}_{cf}$. In addition, a simple power counting (as in Eq.~\ref{T_scaling_of_Pi_f}) shows that the scaling form of $\Pi_f$ leads to $T$--linear resitivities. 
%The use of \eqref{ansatzes} is justified since, for $T\ll \Gamma$, $\mathcal{F}[\phi]$ is approximately minimized by $\phi_{\rm dis}$, the function $\phi$ that minimizes $\mathcal{F}_{\rm dis}[\phi]$, such that
In total, we confirm that
\begin{eqnarray}
\rho \approx \mathcal{F}^{\rm el}_{\rm dis}[\phi^{\rm el}_{\rm dis}] + \mathcal{F}^{\rm el}_{cf}[\phi^{\rm el}_{\rm dis}] = \rho_0 + AT,
\label{variational_resistivity_mfl}
\end{eqnarray}
and similarly $\rho_{\rm th} = \rho_{\rm th,0} + BT$. Importantly, Eq.~\eqref{variational_resistivity_mfl} captures the physical leading low-$T$ behavior, rather than serving as an upper bound. Indeed, consider leading correction to $\phi^{\rm el}_{\rm dis}$, $\delta \phi^{\rm el}$, such that the minimizer of the resistivity functional is $\phi^{\rm el} = \phi^{\rm el}_{\rm dis} + \delta \phi^{\rm el}$. Then, by expanding the right-hand-side of Eq.~\eqref{variational_resistivity_mfl} in $\delta \phi^{\rm el}$ and using the fact that $\phi^{\rm el}_{\rm dis}$ minimizes $\mathcal{F}^{\rm el}_{\rm dis}[\phi^{\rm el}]$, we obtain that the contribution due to $\delta \phi^{\rm el}$ is subleading.
%, which implies that the low-$T$ deviation from the WF law follows \eqref{MFL_WF_deviation}. 
In practice, $\phi_{\rm dis}$ is expected to minimizes the full functional $\mathcal{F}$ in the case of a local $cf$-interaction.

%Note that this is essentially the same reasoning used in the FL case \cite{Ziman_2001} (see also App.~\ref{HS_th_cond}), with the usual Fermi distribution replaced by the generalized one. 

%By virtue of the variational formulation, we can now explicitly see that the validity of the WF law at $T \to 0$ follows from the dominance of the $T$-independent elastic scattering contribution, $\mathcal{F}_{\rm dis}$, over the inelastic term, $\mathcal{F}_{cf}$. 
So far, the QBE provided a simple perspective for the validity of the WF law in the case of local $cf$-interactions, i.e., $\Upsilon_{\boldsymbol{r},\boldsymbol{r}'} = \delta_{\boldsymbol{r},\boldsymbol{r}'}$ (such that $\widetilde{\Upsilon}=1$), for which $\Pi_f$ is completely uniform in momentum space. But in fact, since the key ingredient in our derivation was the PK reduction scheme, the above discussion can be generalized to a class of deformed models with spatially extended $cf$-interactions. See the main text (MFL section) for an example. Indeed, under such deformations, the scattering off of $f$-fermions obtains a momentum dependence, $\Pi_f\left(\nu \right) \to \Pi_f\left(\boldsymbol{q},\nu \right) \equiv \widetilde{\Upsilon}\left(\boldsymbol{q}\right)\Pi_f(\nu)$. However, as long as this momentum dependence is not singular, namely, it can be written as $\widetilde{\Upsilon}\left(\boldsymbol{q}\right) \sim 1 + \eta h(\boldsymbol{q})$ with a sufficiently small $\eta$ and smooth $h$, the low-energy MFL form of the self-energy of the $c$-fermions does not change (as demonstrated above). And most importantly, since the PK reduction scheme is valid, we can repeat the analysis above. It thus follows that this class of deformations, and in particular the example presented in the main text (MFL section), also obeys Eq.~(5) in the main text. 

For completeness, let us note that the main effect of spatially extended $cf$-interactions is to change the nonuniversal prefactor of the leading $-T$ term in $L(T) - L_0$. That is, it modifies the coefficients $A$ and $B$ of the resistivities above.  For the electrical resistivity, this is due to the suppressed contribution of small angle scattering events. This can be seen explicitly by noticing that, in this case, $\phi$ is frequency independent [see Eq. \eqref{ansatzes}] and the momentum and frequency integrations in Eq.~\eqref{factor of 2} factorize such that 
\begin{eqnarray}
    \left\langle \phi,\mathcal{P}_{cf}\phi\right\rangle =\int_{\hat{\boldsymbol{k}},\hat{\boldsymbol{k}'},\boldsymbol{q}}\tilde{\Upsilon}\left(\boldsymbol{q}\right)\left(\phi\left(\hat{\boldsymbol{k}}\right)-\phi\left(\hat{\boldsymbol{k}'}\right)\right)^{2}\delta\left(k_{F}\hat{\boldsymbol{k}'}-k_{F}\hat{\boldsymbol{k}}-\boldsymbol{q}\right)\times (\text{frequency integrals}).
    \label{scaling in Lxx}
\end{eqnarray}
For $\Upsilon = 1$ we retrieve the familiar $1-\cos\theta_{\vec{k},\vec{k'}}$ weighting factor,  while any momentum dependence in $\Upsilon$ will change the weighting and hence the overall prefactor. The $T$-scaling is determined by the frequency intergrals and is hence unaffected. Similar, yet more involved, consideration can be applied to the thermal resistivity \cite{Ziman_2001}.

The validity of PK reduction was the key ingredient to the derivation above, namely, the same analysis can be applied to generic weakly disordered MFLs (and NFLs), provided that the momentum-dependence of the inelastic scattering rate is sufficiently weak (which enables the PK reduction). Furthermore, this means that the low-$T$ deviation from the WF law (Eq.~(5) in the main text) is not a fine-tuned feature of our model, and could serve as a generic criterion for strangeness, as claimed in the main text. 

Lastly, notice that by spatially extending that range of $\Upsilon$ we reduce the characteristic momentum transfer to a narrower region in the Brillouin zone. Denoting the radius of this region by $q_*$, the validity condition for the PK reduction can be roughly estimated as $v_F q_* \gg \Gamma$. It is interesting that by further extending the range of $\Upsilon$, such that $v_F q_* \sim \Gamma$, we effectively invalidate the PK reduction scheme. 
%While intuitively we expect the WF law to hold at the lowest temperatures for the same reasons as above. 
The PK reduction scheme cannot be consistently used in this case. 
%to demonstrate it. 
This scenario is reminiscent of the MFL model studied in \cite{guo_large_2022}.

\subsection{Generalization to the transverse Lorenz ratio}

We obtain the transverse conductivities by a solving the QBE in the presence of a magnetic field for a local $cf-$interaction (introducing extended $cf-$interaction does not change the qualitative physical picture). The reason for this alternative derivation is that the variational principal is not directly valid in the presence of a magnetic field since the operators $P_a$ are not self-adjoint \cite{Ziman_2001}. 
%(a modified variational principal can be formulated to the longitudinal conductivities [Maximum variational principle for conduction problems in a magnetic field])
Physically, this is related to the fact that time-reversal symmetry is broken such that the probability of a scattering of an excitation $(\hat{\vec{k}},\omega)$ to an excitation $(\hat{\vec{k}'},\omega')$ is not equal to the probability in the reverse direction. Nevertheless, the direct solution of the QBE is shows that the transverse Lorenz ratio follows the same $T-$scaling is the longitudinal one provided that the PK reduction holds.

The introduction of a magnetic field follows the same steps as in the case of an electric field. Specifically, we introduce an electromagnetic vector potential via minimal coupling, $\varepsilon_{\vec{k}} \to \varepsilon_{\vec{k+A}}$ where $\vec{A} = \vec{A}_1 + \vec{A}_2$ such that $-d\vec{A}_1/dt = \vec{E}$ and $\nabla \times \vec{A}_2 = \vec{B}$. Then, the QBE is obtained in a similar fashion to the derivation above, where the introduction of electromangetic fields is done via a change of the COM coordinates $\vec{k} \mapsto \vec{k} + \vec{A}$; see e.g. \cite{RammerQFT}. The QBE then takes the form
\begin{eqnarray}
\mathcal{\widetilde{D}}f_{c}\left(\hat{\boldsymbol{k}},\omega,\boldsymbol{r},t\right)=\mathcal{I}_{{\rm coll}}\left[f_{c}\left(\hat{\boldsymbol{k}},\omega,\boldsymbol{r},t\right) \right]
\label{QBE before PK}
\end{eqnarray} 
where $\widetilde{\mathcal{D}} = \mathcal{D} + \nabla_{\boldsymbol{k}_{F}}\varepsilon_{\boldsymbol{k}}\cdot\boldsymbol{E}\partial_{\omega}+\left(\nabla_{\boldsymbol{k}_{F}}\varepsilon_{\boldsymbol{k}}\times\boldsymbol{B}\right)\cdot\nabla_{\boldsymbol{k}_{F}}$. Similarly to the conventional Boltzmann equation, the magnetic part of the Lorentz force nullifies the equilibrium distribution and acts nontrivially only on the non-equilibrium piece. 

To proceed, we insert the parametrization of the full distribution, $f_c \equiv f_0 + \delta f$, into $\mathcal{I}_{\rm coll}$. Then, we use (a) the fact that by definition $\mathcal{I}_{\rm coll}\left[f_0\right] = 0$ \cite{kamenev_field_2011}, and (b) the fact that $\delta f (-\hat{\vec{k}}) = - \delta f (\hat{\vec{k}})$, which implies that the terms proportional to $\delta f(\hat{\vec{k}}')$ vanish in the integration over $\hat{\vec{k}}'$. It is then straightforward to express the collision integral in terms of the self-energy of the $c-$fermions:
\begin{eqnarray}
    \mathcal{I}_{\rm coll}[f] = 2 \Sigma''_R(\omega) \delta f.
\end{eqnarray}
Finally the QBE for in the presence of an electric and magnetic field is given by 
\begin{eqnarray}   v_{F}\hat{\boldsymbol{k}}\cdot\boldsymbol{E}\partial_{\omega}f_{0}\left(\omega\right)+v_{F}\left(\hat{\boldsymbol{k}}\times\boldsymbol{B}\right)\cdot\nabla_{\boldsymbol{k}_{F}}\delta f\left(\hat{\boldsymbol{k}},\omega\right)=2\delta f\left(\hat{\boldsymbol{k}},\omega\right)\Sigma_{R}^{''}\left(\omega\right).
\label{QBE form with magnetic field}
\end{eqnarray}
The QBE in the presence of a small thermal gradient is identical to the above with the replacement $\vec{E} \to \beta \omega \nabla_{\vec{r}} T$. The solution of the QBE is obtained by inserting the ansatz \cite{Ziman_2001} ${\delta f}\left(\hat{\boldsymbol{k}},\omega\right)=  k_F \hat{\vec{k}} \cdot \vec{\delta f} \left(\omega\right)$, which yields (in agreement with \cite{patel_magnetotransport_2018})
\begin{eqnarray}
\delta f_{i}\left(\omega\right)=\frac{v_{F}}{k_{F}}\partial_{\omega}f_{0}\left(\omega\right)\left(2\Sigma''_{R}\left(\omega\right)\delta_{ij}+\epsilon_{ij}B\frac{v_{F}}{k_{F}}\right)^{-1}F_{j}
\end{eqnarray}
with $F_j = E_j, \beta \omega \partial_j T$ for an applied electric field and thermal gradient, respectively (the magnetic field is assumed to point along $\hat{z}$, i.e. out of the plane), and $\epsilon_{ij}$ is the antisymmetric tensor in two dimensions. We recall that, in our model, by definition the (expectation values of the) electrical and thermal currents (per flavor) are given by 
\begin{eqnarray}
J_{\text{el},i}=-\nu_{0}\int_{\hat{\boldsymbol{k}},\omega}\vec{v}_{F}\delta f\left(\hat{\boldsymbol{k}},\omega\right)
\end{eqnarray}
and 
\begin{eqnarray}
J_{\text{th},i}=-\nu_{0}\int_{\hat{\boldsymbol{k}},\omega}\vec{v}_{F}\omega\delta f\left(\hat{\boldsymbol{k}},\omega\right).
\end{eqnarray}
We can therefore obtain the transverse electrical and thermal conductivities,
\begin{eqnarray}
    \sigma_{xy}=\frac{v_{F}^{2}\nu_{0}}{16T}\int\frac{d\epsilon}{2\pi}\text{sech}^{2}\left(\frac{\epsilon}{2T}\right)\frac{\omega_c/2}{\Sigma_R''\left(\epsilon\right)^{2}+(\omega_c/2)^{2}}
    \label{sigma xy}
\end{eqnarray}
and
\begin{eqnarray}
    \kappa_{xy}=\frac{v_{F}^{2}\nu_{0}}{16T^{2}}\int\frac{d\epsilon}{2\pi}\text{sech}^{2}\left(\frac{\epsilon}{2T}\right)\epsilon^{2}\frac{\omega_c/2}{\Sigma_R''\left(\epsilon\right)^{2}+(\omega_c/2)^{2}}, 
    \label{kappa xy}
\end{eqnarray}
where the cyclotron frequency is given by $\omega_c=\left(v_{F}/k_{F}\right)B$. 

We can now explicitly see that $L_{xy} - L_0 \propto - T$ in agreement with Eq.~(5) in the main text. For example, in the simplest case where the magnetic field is sufficiently small ($\omega_c \ll \Gamma$), the conductivities obey the relation $\alpha_{xy}(T) = (\omega_c/2\Gamma) \alpha_{xx}(T)$ ($\alpha = \sigma,\kappa$) to leading order in $\omega_c$, which automatically guarantees the desired behavior. Note also that even when $\Gamma \to 0$ the WF law is obeyed ($\omega_c$ takes the role of the disorder term) but with a deviation that scales as $T^2$. 

Analogously to the longitudinal case, the leading $T$-scaling of the transport rates is governed by the form of ${\rm Im}\Pi^R_f(\nu)$ in $\mathcal{I}_{cf}$, which is unaffected by the introduction of extended interactions. Hence, as in the longitudinal case, spatially extended $cf$-interactions changes the prefactor of the $-T$ term in $L_{xy}(T)-L_0$, but not its scaling form. More generally, the $T$-scaling remains unchanged as long as the inelastic scattering mechanism has sufficiently weak momentum dependence. This is exactly the validity condition of the PK reduction. We thus observe that, similarly to the discussion on the longitudinal Lorenz ratio, our conclusion holds for models of weakly disordered MFLs (of NFLs) where the PK reduction scheme can be applied. In other words, the leading deviation of the transverse Lorenz ratio satisfies the same generic behavior as the longitudinal Lorenz ratio, provided that the PK reduction is valid.

\end{document}